\documentstyle[aaspp4,fleqn,11pt,psfig]{article}
\newcommand{\siml}{\stackrel{<}{\sim}}

\newcommand{\dt}[1]{{{{\rm d}#1}\over{{\rm d}t}}}
\newcommand{\dr}[1]{{{{\rm d}#1}\over{{\rm d}r}}}
\newcommand{\deta}[1]{{{{\rm d}#1}\over{{\rm d}\eta}}}
\newcommand{\ddr}[1]{{{\rm d}\over{{\rm d}r}}#1}

\newcommand{\ddlnT}[1]{{{\rm d}\over{{\rm d}\ln T}}#1}
\newcommand{\ddlnr}[1]{{{\rm d #1}\over{{\rm d}\ln r}}}
\newcommand{\dDlnr}[1]{{{\rm d}\over{{\rm d}\ln r}}#1}

\newcommand{\dDmr}[1]{{{\rm d}\over{{\rm d}M_r}}#1}
\newcommand{\Dt}[1]{{{{\rm D} #1}\over{{\rm D}t}}}
\newcommand{\DDt}[1]{{{\rm D}\over{{\rm D}t}}#1}

\newcommand{\pMr}[1]{{{\partial #1}\over{\partial M_r}}}

\newcommand{\Plnr}[1]{{{\partial}\over{\partial \ln r}}#1}

\newcommand{\bq}{\begin{equation}}
\newcommand{\eq}{\end{equation}}
\newcommand{\bqa}{\begin{eqnarray}}
\newcommand{\eqaa}{\end{eqnarray*}}
\newcommand{\bqaa}{\begin{eqnarray*}}
\newcommand{\eqa}{\end{eqnarray}}

\newcommand{\gge}{g^{ik}}
\newcommand{\ggf}{g^{jk}}

\newcommand{\brho}{\bar{\rho}}
\newcommand{\bPhi}{\bar{\Phi}}

\newcommand{\bu}{\bar{u}}
\newcommand{\bT}{\bar{T}}

\lefthead{D. R. Xiong, Q. L. Cheng \& L. Deng}
\righthead{Nonlocal Time-dependent Convection Theory}

\begin{document}

\title{Turbulent Convection and Pulsational Stability of Variable Stars\\
II. Oscillations of RR Lyrae and Horizontal Branch  Red Variable Stars}
\author{D. R. Xiong$^1$, Q. L. Cheng$^1$ \& L. Deng$^2$}
\affil{$^1$Purple Mountain Observatory, Academia Sinica, Nanjing 210008, P.R. 
China\\
$^2$Beijing Astronomical Observatory, Academia Sinica, Beijing 100080, P.R. 
China}

\begin{abstract}
Using a nonlocal time-dependent theory of convection, we have calculated 
the linear non-adiabatic oscillations of the Horizontal Branch (HB) 
stars, with both the dynamic and thermodynamic coupling between convection 
and oscillations carefully treated. Turbulent pressure and turbulent viscosity
have been included consistently in our equations of non-adiabatic pulsation. 
When the coupling between convection and oscillations is ignored, for all 
models with $T_e\leq 7350$, the fundamental through the second overtone
are pulsationally unstable; while for $T_e\leq 6200$ all the models are unstable 
up to (at least) the 9th overtone. When the coupling between convection and 
oscillations is included, the RR Lyrae instability strip is very well predicted. 
Within the strip the most models are pulsationally unstable only for the 
fundamental and the first few overtones. The turbulent viscosity is an important 
damping mechanism. Being exclusively distinct from the luminous red variables 
(long period variables), the HB stars to the right of 
the RR strip are pulsationally stable for the fundamental and low-order overtones,
but become unstable for some oen of the high-order overtones. This may provide a 
valuable clue for the short period, low amplitude red variables found outside 
the red edge of the RR strip on the H-R diagram of globular clusters. Moreover,
we present a new radiation modulated excitation mechanism functioning in a 
zone of radiation flux gradient. The effects of nonlocal convection and the 
dynamic coupling between convection and oscillations are discussed. The 
spatial oscillations of the thermal variables in the pulsational calculations 
have been effectively suppressed.

\end{abstract}


\section{Introduction}\label{Sect1}
The main goal of the series of our present work is to study the effects of
convection on the stability of variable stars of different types. Gough (1977) 
and Xiong(1977) have pointed out independently that a nonlocal time-dependent 
theory of convection is needed in order to account for the dynamic coupling 
between convection and oscillations. Based on the general equations of fluid 
dynamics and the theory of turbulence, we have developed such a nonlocal 
time-dependent statistical theory of convection (Xiong, Cheng \& Deng 1997), 
which gives an equation for turbulent viscosity that is formally similar to 
the Stokes equation for viscous gas. The theoretical reasonings and the 
correct expression of the gradient approximation for the triple correlations 
(the 3rd-order correlation functions of the turbulent fluctuations of velocity 
and temperature) are provided by the dynamic equations. We have presented 
results of the linear nonadiabatic oscillations for long period variables in 
the first paper of this series (Xiong, Deng \& Cheng, 1997, hereafter Paper I),
in which a so-called Mira instability region outside the Cepheid instability 
strip has been defined. In this paper, we will apply our theory to the study 
of the pulsational stability of HB stars.
\par
The theoretical progress of RR Lyrae star research is due to many
authors, among which we recall Christy (1966), Iben (1971), 
Baker and Gough (1979), Deupree (1977a,b,c), Xiong (1980), Stellingwerf (1984), 
Bono and Stellingwerf (1993), Guzik and Cox (1993) and Bono et al (1995, 1997).
Christy(1966) and Iben(1971) completely ignored the coupling between convection
and oscillations, therefore they gave no definition for the red edge of
the RR instability strip. Baker \& Gough (1979) followed Gough's local
time-dependent mixing length theory of convection (Gough 1977), while 
Xiong (1980) used a local time-dependent statistical theory of convection
(Xiong 1977). The later two papers gave very close results, and both
reached a definite red edge of the RR instability strip. The detailed
comparison of the two was given by Balmforth \& Gough (1987).
The coupling between convection and oscillations was attempted by
Stellingwerf (1984) and Bono et al (1993, 1995, 1997) using a rather
simplified nonlocal time-dependent theory of convection.
In this paper, we have put great emphasis on the theoretical
baseline of the convection theory, using which we have calculated
the linear non-adiabatic pulsation for RR Lyrae stars. We are going
to show the advantages of the present theory over the local
time-dependent theory of convection.
\par
It is claimed that a bunch of short period, small amplitude variables
were found outside the RR instability strip on the H-R diagram of
some globular clusters (Yao 1981, 1986, 1987; Yao et al. 1993a,b,
1994). In this paper we try to give a theoretical interpretation 
of such variables. In the following section, the main results of
the calculated models for non-adiabatic pulsation are presented. We also
proposed a new excitation mechanism, i.e. the so called
``radiation modulated excitation mechanism'',
which works in the region of radiation flux gradient.
We have explained in Section~\ref{sec3}
the reason for the spatial oscillations of
the thermodynamic quantities that happen in the numerical
computations of stellar pulsation . A concise discussion and conclusions 
of our work are in Section~\ref{sec4}

\section{The Numerical Results}\label{sec2}
We have calculated the fundamental through 11th order overtone
modes of linear non-adiabatic oscillations for  six series of HB star 
models with a intermediate metal abundance.  All the model series have the 
same chemical composition: $(X,Z)=(0.76,0.0004)$.  Their mass, luminosity, 
chemical composition and the characteristies of pulsational instablity region 
are listed in Table 1. The model parameters are: 
$M=(0.65-0.80)M_\odot$; $L=(1.5-2.2) \times 10^{35}erg/s$ 
($M_{bol}=0.77-0.35$); and effective temperature in the range of 
$5150$ -- $8075K$, where the coolest model is very close to the
Hayashi track; the convection parameters $c_1=c_2=0.75 - 1.00$,
which correspond to the same energy transport efficiency 
as taking $\alpha=l/H_p\approx 1.5 - 2.0$ in the original Vitense
theory, and the e-folding length of convective overshooting 
of about (1.05 - 1.4)$H_p$. The coupling between convection and
oscillations is carefully treated by using our nonlocal
time-dependent theory of convection (Xiong, 1989, Xiong, Cheng \&
Deng 1997). All the mathematical and numerical schemes are
the same as in Paper I to which we refer for all the details
in the equations and boundary conditions.
\par
The lower boundary is set in the convective overshooting zone 
($T= 8\times 10^{6}K$), and the upper one is set optically thin enough in
the stellar atmosphere (optical depth $\tau=0.01$).  
We use a simplified MHD equation of state (Hummer and Mihalas, 1988;
Mihalas et al. 1988; Dappen et al. 1988) in the present work. 
An analytic approach to the OPAL tabular opacities (Rogers and Iglesias 1992) 
and the low-temperature tabular opacities (Alexander 1975) is used for the 
calculation of opacity. The Henyey method (Henyey et al. 1964) 
is used for numerical calculations. The envelope model of stars 
is divided into 800 zones with approximately equal intervals in $ln ~~T$.
However, the mesh points have been increased in the stellar atmosphere,
where the temperature gradient is very small. The variations of the 
temperature and pressure within any mesh zone are less than 0.03 dex. 
These mesh points have enough spatial resolution for the highest overtone 
in the present work. The eigen-functions of the 10th overtone for a red 
HB model are plotted in Fig.~\ref{fig1}.

\subsection{Radiative Modulation Excitation}\label{subsec2.1}
In order to demonstrate the effects of convection on the stability of stellar 
pulsation, we have considered separately two cases, i.e. including and 
ignoring the coupling between convection and oscillations (however, 
convection has been included in the calculation of equilibrium models). 
The treatment of decoupling can be referred to Paper 1. Simply saying, 
it is by setting the convective variables to zero. Table~2 gives the 
linear amplitude growth rates for the first 12 eigen-modes for the case 
that the coupling between convection and oscillations is ignored. The 
first column is the serial number of the model, the second is the 
effective temperature $T_e$, 3rd -- 14th columns are respectively the 
amplitude growth rates for the fundamental to the eleventh overtone.
As it is clear from Table~2,  for all models with $T_e\leq 7325$, 
the fundamental through the second overtone are pulsationally unstable; 
while all the models with $T_e\leq 6200$ K are unstable up to (at least) 
the 9th overtone. This makes it clear that the red edge of the RR 
instability strip cannot be determined when the coupling between 
convection and oscillations is ignored. Fig.~\ref{fig2}a-f depicts the
variation of the integrated work versus the depth ($\log T$) for six 
HB stellar models with different effective temperatures. 
\par
It can be seen from Fig.~\ref{fig2} that the excitation of pulsationmainly 
comes mainly from 3 regions, namely, the $\kappa$ mechanism (KM) which works in
the vicinity of the helium and hydrogen ionization regions, and the 
modulation excitation mechanism working in the gradient zones of 
radiation flux (RGZ) on the top and at the bottom of convection zone,
about which we are going to talk in details below. For the HB
stars with $T_e\leq 5800K$, the ionization region of helium is almost 
completely convective and the radiation energy transport is nearly 
negligible ($L_r/L<1\%$). Therefore, the KM working
near the hydrogen and helium ionization regions is not at all
important. The pulsation is excited by the KM  
and the radiation modulated excitation (RME) mechanism above the 
hydrogen ionization region, where $dW_p/dlog T<0$ (Fig.~\ref{fig2}a,b).
There is a lot of work concentrated on the KM in the 
literature. We will focus on the new mechanism of pulsational excitation, 
i.e. the modulation excitation that works in the RGZ on the top and at the 
bottom of the convective zone. In such RGZ the radiation flux will be 
modulated when the star pulsates. The function of such modulation is to 
gain energy from the surroundings through radiation while the fluid
element is hot, and to loss energy while it is cool. Thus
on the P-V work plot, a positive Carnot cycle forms,
that is to say, radiation energy is converted
into pulsational kinetic energy by such a process.
In the following, we are going to explain how such an
excitation mechanism works by using the integrated
work. Following our previous work (Cheng \& Xiong 1997),
the normalized integrated work can be expressed as

\bq
W_P=-{\pi \over{2E_k}}\int^{M_0}_{M_b}{1\over\rho^2}I_m
    \left(\delta P\delta\rho^*\right)dM_r,
\label{Eq1}
\eq
where, $E_k$ is the total kinetic energy of oscillations,

\bq
E_k={1\over 2}\omega_r^2\int^{M_0}_{M_b}\delta r\delta r^*dM_r,
\label{Eq2}
\eq
Starting from the equation of conservation of energy, one
can easily have,

\bq
\dt{P}-\Gamma_1{P\over\rho}\dt{\rho}=
-\left(\Gamma_3-1\right)\rho\dDmr{\left( L_r+L_c+L_t\right)} ,
\label{Eq3}
\eq
where $L_r$, $L_c$ and $L_t$ are respectively the 
corresponding luminosities of the radiative, 
convective and turbulent kinetic energy fluxes.
Inside the equilibrium model of stellar envelope,

\bq
L_r+L_c+L_t=L=const.
\label{Eq4}
\eq
Linearizing Eq.~(\ref{Eq3}), we have

\bq
\delta P=\Gamma_1{P\over\rho}\delta\rho-\left(\Gamma_3-1\right)
{\rho\over{i\omega}}\dDmr{\left(\delta L_r+\delta L_c+\delta L_t\right)},
\label{Eq5}
\eq
Inserting Eq.~(\ref{Eq5}) into Eq.~(\ref{Eq1}), we have

\bq
W_P=-{\pi\over{2\omega_rE_k}}\int^{M_0}_{M_b}
R_e\left\{\left(\Gamma_3-1\right){{\delta\rho^*}\over\rho}
\dDmr{\left(\delta L_r+\delta L_c+\delta L_t\right)}\right\}dM_r.
\label{Eq6}
\eq
\par
For not losing any generality and for the sake of convenience in 
discussing the coupling between convection and oscillations, we have kept in 
Eqs.~(\ref{Eq5}) and (\ref{Eq6}) the terms related with convection, i.e. 
$\delta L_c$ and $\delta L_t$. If $\delta L_c$ and $\delta L_t$ are set to 
zero, this will lead to exclusion of the coupling between convection and 
oscillations. We note

\bq
{{\delta L_r}\over L_r}=\ddlnT{\left({{\delta T}\over T}\right)}
         +\left( 4-\chi_{_T}\right){{\delta T}\over T}
         -\chi_{_P}{{\delta P}\over P}+4{{\delta r}\over r},\\ 
\label{Eq7}
\eq
\bq
{{\delta L_c}\over L_c}=\left( \delta+C_{p,P}\right){{\delta P}\over P}
         +\left( 1+C_{p,T}-\alpha\right){{\delta T}\over T}
         +2{{\delta r}\over r}+{{\delta V}\over V},\\
\label{Eq8}
\eq
\bq
{{\delta L_t}\over L_t}={{\delta L_1}\over L_1},\\
\label{Eq9}
\eq
\bq
\left(\Gamma_3-1\right){{\delta\rho^*}\over\rho}\approx {{\delta T^*}\over T},\\
\label{Eq10}
\eq
\bq
{{\delta P}\over P}\approx {\Gamma_2\over{\Gamma_2-1}}{{\delta T^*}\over T},
\label{Eq11}
\eq
where $\alpha$ and $\delta$ are the coefficients of expansion and compression of 
gas.  Putting Eqs.~(\ref{Eq7}) - (\ref{Eq11}) into Eq.~(\ref{Eq6}), Wp can be 
expressed as a sum of the following five terms

\bq
W_P=W_{KM}+W_{RME}+W_{CE}+W_{CME}+W_{L_t},
\label{Eq12}
\eq
where $W_{KM}$ and $W_{RME}$ are the contributions of the nomal KM 
and RME respectively, 

\bq
W_{KM}={\pi\over{2\omega_rE_k}}
         \int^{R_0}_{r_b}R_e\left\{{{\delta T^*}\over T}\ddr\left[
        \left(\chi_{_T}+{\Gamma_2\over{\Gamma_2-1}}\chi_{_P}
        -4\right){{\delta T}\over T}
        -\ddlnT\left({{\delta T}\over T}\right)-4{{\delta r}\over r}
        \right]L_r\right\} dr, 
\label{Eq12.1}
\eq

\bq
 W_{RME}={\pi\over{2\omega_rE_k}}
\int^{R_0}_{r_b}R_e\left \{ {{\delta T^*}\over T}\left[
        \left(\chi_{_T}+
        {\Gamma_2\over{\Gamma_2-1}}\chi_{_P}-4\right){{\delta T}\over T}
        -\ddlnT{\left({{\delta T}\over T}\right)}-4{{\delta r}\over r}\right]
        \dr{L_r}\right \} dr,
\label{Eq12.2}
\eq
$W_{CE}$ and $W_{CME}$ are the contributions of the convective thermal flux and 
the convective (flux) modulation excitation (CME),

\bq
W_{CE}=-{\pi\over{2\omega_rE_k}}
       \int^{R_0}_{r_b}R_e\left \{ {{\delta T^*}\over T}\ddr\left[
        \left( 1+C_{p,T}-\alpha+{\Gamma_2\over{\Gamma_2-1}}
        \left( \delta+C_{p,P}\right)\right){{\delta T}\over T}
        +2{{\delta r}\over r}+{{\delta V}\over V}\right]L_c\right \} dr,
\label{Eq12.3}
\eq

\bq
W_{CME}=-{\pi\over{2\omega_rE_k}}
       \int^{R_0}_{r_b}R_e\left \{ {{\delta T^*}\over T}\left[
        \left( 1+C_{p,T}-\alpha+{\Gamma_2\over{\Gamma_2-1}}
        \left( \delta+C_{p,P}\right)\right){{\delta T}\over T}
        +2{{\delta r}\over r}+{{\delta V}\over V}\right]\dr{L_c}\right \}dr,
\label{Eq12.4}
\eq
and $W_{L_t}$ is the contribution of the turbulent kinetic energy flux,
\bq
W_{L_t}=-{3\pi\over{2\omega_rE_k}}
       \int^{R_0}_{r_b}R_e\left \{ {{\delta T^*}\over T}\left[
        L_1\ddr\left( {{\delta L_1}\over L_1}\right)+{{\delta L_1}\over L_1}
        \dr{L_1}\right] \right \} dr.
\label{Eq12.5}
\eq
In our nonlocal convection theory $|L_1/L|\ll 1$, so $W_{L_t}$ is negligible in 
comparision with $W_{KM}$, $W_{RME}$, $W_{CE}$ and $W_{CME}$.
\par
When 
${\rm d}\left\{ \left[\chi_{_T}+\chi_{_P}\Gamma_2/(\Gamma_2-1)-4\right]|\delta T/T|\right\}/{\rm d}r>0$ 
($<0$), KM plays a destability (stability) effect for stellar oscillations.
The destability effect is enhanced in the hydrogen and helium ionization 
regions, because $\Gamma_2$ is small due to the ionization of hydrogen and helium 
(Gamma mechanism). When the coupling between convection and 
oscillations is ignored, $W_{CE}=W_{CME}=0$ and $W_P=W_{KM}+W_{RME}$. 
This theoretical prediction has been justified by our numerical calculations. 
\par
The integrated works $W_{KM}$ and $W_{RME}$ are also plotted in Fig.~\ref{fig2}. 
It can be seen from Fig.~\ref{fig2} that in the stellar interior, where the 
pulsational amplitudes increase outwards, the $W_{KM}$ increases (decreases) 
with increase (decrease) of $\kappa_T$. 

There is a radiative damping region ($dW_P/dLog T>0$) in the deep radiative 
interior, where $W_{KM}<0$. For stars with very low effective temperature, 
such a radiative region goes even deeper into the stellar interior where the 
pulsational amplitude is small, and the radiative damping is small as well, 
as this is seen in Fig.~\ref{fig2}a,b. As the temperature increases, the 
envelope convective zone retreats towards the surface. The radiative damping 
region progresses outwards, and the radiative damping increases 
(Fig.~\ref{fig2}c-f). There is also a radiative damping region ($dW_{KM}/dLog T>0$) 
in the stellar atmosphere, because the pulsational amplitude of $\delta T/T$ 
decreases towards surface. 
\par
The RME exists only in the RGZ. In a stellar envelope RGZ is present only at the 
top and the bottom of the convective zone [Up Radiative Gradient Zone (URGZ) and 
Bottom Radiative Gradient Zone (BRGZ)]. In the BRGZ where ${\rm d}L_r/{\rm d}r<0$ 
and $\chi_{_T}+\chi_{_P}\Gamma_2/\left(\Gamma_2-1\right)-4<0$, $W_{RME}$ 
is positive. Therefore, the BRGZ is 
an excitation region of oscillations (Fig.~\ref{fig2}b-e). In the URGZ the above 
two inequalities are both reversed. This is another excitation region (Fig.~\ref{fig2}a-d). 
Such an excitation mechanism is related to opacity $\kappa$, but it is distinct from 
the normal KM. The normal KM can work in 
purely radiative stellar configuration, for example, such as the excitation of 
$\beta$ Cephei variables, which have almost no convection zone. This new 
exciting mechanism, however, works only in the RGZ. We will call it the 
radiation modulated excitation (RME), so as to distinguish it from the 
normal KM.
\par
The location of the RME at the top of a convective zone overlaps with that of 
the normal KM, so it is difficult to distinguish them in the total 
power diagram ($W_P-log T$). But, they are clearly distinguished with the 
curves of $W_{KM}$ and $W_{RME}$. The location and the efficiency of the RME at 
the bottom of the convective zone depend very sensitively on the effective 
temperature of 
the stellar model under consideration. For extremely low temperature models, 
the envelope convective zone is very extended, and this leads to a very deep 
modulated excitation region at the bottom of the convective zone. However, the 
pulsational amplitude in this region is very small,  so a very weak 
excitation is expected. The main excitation comes from the KM
and RME mechanisms in the URGZ (Fig.~\ref{fig2}a,b) for the cool HB 
stars. In Fig.~\ref{fig2}b, the bottom RME is only barely visible. 
As the effective temperature increases, the bottom RME raises up together with 
the shrinking of the convective zone (Fig.~\ref{fig2}a-d). 
The effect of bottom RME is clearly illustrated in Fig~\ref{fig2}b and 
Fig~\ref{fig2}c. The bottom RME starts to work at $\log T\approx 5.7$ and 5.4
respectively, corresponding to the location of the bottom of the convective 
zone and being much lower than the second ionization region of helium where the 
normal KM works. For a HB star with $T_e=5900$ the bottom RME 
region is connected or overlaps with the second ionization region of helium 
(Fig.~\ref{fig2}c). Numerically, RME is stronger than the KM in 
the second ionization region of helium. For models with $T_e=6200K$ the bottom 
RME region is near the second ionization region of helium (Fig.~\ref{fig2}d). 
From the total power diagram ($W_P-log T$) it is difficult to distinguish the 
RME from the KM. The effects of RME still can be seen from the large increase 
of integrated work $W_{RME}$ near $log T=4.8$ corresponding to the location of 
the bottom of the convective zone.
\par
For the blue HB stars , the convective energy transport 
is insignificant since the convective envelope is very shallow, 
and the KM functioning in the second 
ionization region of helium becomes dominant, while RME is almost not at all 
working. We can also see from Fig.~\ref{fig2}f that the excitation due to the 
KM cannot balance the radiative damping in deeper layers for 
high enough effective temperature models and the star becomes 
pulsationally stable. This is the reason of the existence of the blue 
edge of the RR instability strip.
\par
One can see from Fig.~\ref{fig2} that there is a bump near $log T\approx 5.3$.
It results from the bump of the new OPAL opacity.

\subsection{The RR Instability Strip}\label{subsec2.2}
As we have repeatedly stressed, when the coupling between convection and oscillations
is ignored, all the HB star models with $T_e\leq 7550K$ are pulsationally
unstable. Therefore, no definition for the red edge of RR strip can be given.
However, when the coupling is taken into consideration, the situation
becomes very promising. We have listed in Table~3 the linear amplitude growth 
rates of pulsation (per period) for the first twelve modes. From this table we 
find that when the coupling between convection and oscillations is considered, 
the fundamental and low-order overtones ($n\leq 3$) become pulsationally stable for
all the HB stellar models with $T_e\leq 5940$. Namely, there is an apparent red 
edge of the RR instability strip. The amplitude growth rates 
of the fundamental and the first overtone versus the effective temperatures for 
the model series 2 - 4 are plotted in Fig.~\ref{fig3}. In Fig.~\ref{fig4} the 
integrated works versus depth are given for the fundamental modes of 6 HB 
stellar models, where $W_P$, $W_{Pt}$ and $W_{vis}$ are, respectively, the 
contributions due to the gas (including radiation), turbulent pressure and 
turbulent viscosity, while $W_{all}=W_P+W_{Pt}+W_{vis}$ is the total 
integrated work, the numerical value of which at the surface
is approximately the same as the linear pulsational
amplitude growth rate $\eta=-2\pi\omega_i/\omega_r$
computed with the non-adiabatic scheme.
Here $\omega_i$ and $\omega_r$ are the
imaginary and real part of the complex pulsational angular frequency 
($\omega=i\omega_i+\omega_r$).
For the real definitions of $W_P$, $W_{Pt}$ and $W_{vis}$ we refer to our
previous work (Cheng \& Xiong 1997). Here we only recall that of $W_{vis}$
to make our discussion clearer. $-W_{vis}$ is the random turbulent
kinetic energy converted from the pulsational energy due to turbulent 
viscosity, which is finally turned into thermal energy through molecular 
viscous processes. $W_{vis}$ is always negative. Therefore, This implies 
a damping mechanism. 
\par
We can see from Fig.~\ref{fig4} that, for HB models with $T_e\geq 6000K$, the 
radiation energy transport is becoming dominant gradually in their second 
ionization region of helium. For these stars, the main source of excitation is 
the KM in the second ionization region. When the 
effective temperature drops below $5900K$ for the HB models, the hydrogen and  
helium ionization regions are fully convective, and the radiation contributes 
very little to the total energy flux ($L_r/L\leq 1\%$). Therefore, the 
KM is almost inhibited. The convection becomes the dominant 
factor for determination of the pulsational stability of stars. Thus, the low 
temperature HB stars with $T_e\leq 5940K$  become pulsationally stable. The
coupling between convection and oscillations is the cause  why there exists a 
red edge for the RR Lyrae instability strip. Fig~\ref{fig6} plots the 
variations of temperature for the lower boundaries of the convective zone and 
the ionization regions of hydrogen and helium versus the effective temperature 
of the HB models under consideration. As demonstrated in Fig.~\ref{fig6}, the 
red edge of the RR Lyrae instability strip is defined when the bottom of the 
convective zone is just extended slightly below the second ionization region 
of helium. 

\subsection{The Short Period Red Variables Beyond the RR Lyrae Strip}
\label{subsec2.3}
Yao and his collaborators found some short period red variables
outside the RR Lyrae strip in the H-R diagrams of a few globular clusters
(Yao 1981, 1986, 1987; Yao et al 1993a,b, 1994). The light amplitude of such 
red variables is as small as a few percent of a magnitude. These stars are 
usually multi-periodic or irregular; their typical period or the time-scale of 
variability is a few tenths of a day. Taking the variation of light of such
red variables as a result of intrinsical pulsation, the natural conclusion 
is that the stars all pulsate at high-order overtones. In general, the dwarfs tend 
to be non-radially excited while the giants are radially excited. Therefore, 
we think that the short period red variables are pulsating in high radial 
overtones. Of course, we cannot exclude the possibility that they pulsate at 
low degree non-radial modes. It is very interesting to see from Table~3 that, 
when the coupling between convection and oscillations is considered, all the 
fundamental and low-order overtones of the red HB models for $T_e\leq 5940K$ 
are pulsationally stable; while some high-order overtones ($n\geq 4$)are 
pulsationally unstable. We also show that these low temperature 
red variables possess nearly the same period as that of the variables found 
to right side of the RR Lyrae strip by Yao and his collaborators. 
A general regularity of variable stars is that 
when a star pulsates at high-order overtones, its pulsation amplitude is usually 
very small and tends to be multi-periodic. The results of our theoretical 
modeling are likely to be a reasonable explanation for the red variables found 
outside the RR Lyrae strip in globular clusters  
and for their qualitative properties, though there 
still exist some questions to be studied. According to our theory, it seems that 
all the low temperature HB stars ought to be pulsationally unstable at 
high-order overtones. This raises a naive question: why these short period 
oscillations were observed only for some, not for all the low temperature HB stars?
A straightforward answer to such a question is that they all might be pulsationally
unstable but are not detectable with current instrumental precision. This 
seems to be a reasonable answer. However, there is still a question to be asked: 
what is the drive to make some of the stars pulsate with higher amplitudes? 
We are not ready to give satisfactory explanations to such questions. 

\subsection{Comparison with Other Theoretical Results and Observations}
\label{subsec2.4}
\par
The RR Lyrae instability strip yielded by our numerical computations of the
linear non-adiabatic oscillations is $3.774\leq\log T_e\leq 3.884$ for the 
second model series ($c_1=c_2=0.85$). The location of instability strip of the 
fundamental mode agrees well with Bono's nonlinear theoretical results 
(Bono, et al. 1997). The temperature of the red edge for the fundamental mode is 
slightly higher, while the temperature of the blue edge is slightly lower 
than theirs. Our blue edge of the first overtone  is much hotter than that given 
by Bone's nonliner results. Their blue edge, we think, seems to be too red . Our 
theoretical red edge of the first overtone is slightly hotter than the fundamental 
mode red edge. Obviously, it is too red in comparison with observations. It is 
the intrinsic defect of the linear theory. This is a problem of mode choice. It is 
impossible to decide which mode can finally become an unstable one with limited 
amplitude for the linear theory when several modes are linearly pulsationally unstable. 
\par
For the globular cluster M3 whose metallicity is close to what we adopted in 
this work. Sandage (1990) determined observationally a RR Lyrae instability 
strip $3.791\leq\log T_e\leq 3.880$. This infers that the theoretical result 
matches very well the observed one for the blue edge of the instability strip, 
while the red edge is 240 K lower than the observed one. For the same mass, 
luminosity, effective temeperatute and chemical composition, the convection 
zone deepens as $c_1$ increases. Therefore, the location of the red edge 
of the instability strip can be adjusted by modifying the convective parameter 
$c_1$. We calculated the linear non-adiabatic oscillations of three series 
of HB star models with different $c_1$. The effective temperatures of the blue 
and red edges for the fundamental and the first overtone 
are listed in Table 1. The red edge is still 170 K lower than observation 
even when $c_1$ is increased to 1.0. The energy transport efficiency in this 
case corresponds to that of taking $\alpha=l/H_p\approx 2.0$ in the original 
Vitense theory. It seems impossible that $c_1$ can be much larger than 1. We do 
not know whether this is due to the imperfection of our theory or because the 
temperature of the red edge determined by Sandage is too high.

\subsection{Luminous Red Variables (LRV) and Red HB Stars (RHBS)}
\label{subsec2.5}
\par
Although LRV and RHBS are in the different evolutional statuses, they all 
possess very extended convective envelopes. Convection is the most dominant 
factor controlling their pulsational stability. However, their pulsational 
properties are entirely different: LRV are pulsating at the fundamental 
and the first few overtones, while all higher ($n > 4$) overtones are stable. 
On the contrary, for RHBS the fundamental and lower order overtones ($n \siml 4$) are 
pulsationally stable, while some higher order overtones are unstable. It is hard to 
understand this acute difference at the first glance. In fact, this is because 
these two types of stars have quite different mass-luminosity ratios: 
The masses of LRV do not differ much from those of RHBS, while the luminosities 
of LRV are higher than those of RHBS up to 2 orders of magnitude. Thus their 
mass-luminosity ratios differ about in 2 orders of magnitude. This implies that the 
envelope structure of LRV is very different from that of RHBS. For LRV, the 
central condensation is very high. This will lead to very different 
pulsational properties for these two types:
\begin{enumerate}
\item An important factor is that the thermal time scales of pulsational 
amplitude growth for these two types are very different, and this is due to 
the tremendous difference of mass-luminosity ratios. According to Eq.~(\ref{Eq4}) 
in Paper I, this thermal time scale can be expressed as $4\pi r^3\rho C_p T/L$. 
Since the luminosity of LRV is high and the gaseous density in their envelope 
is very low in comparison with RHBS, the thermal time scale of LRV 
is very short (compared with their pulsation period). It has been shown by actual 
calculations that the amplitude growth rate (the amplitude increment in a 
pulsational period) of LRV is higher than that of RHBS by about 2--3 orders of 
magnitude. This means that, for LRV, the excitation (damping) by radiative 
(and convective) energy transport is very strong. Therefore, for fundamental 
and low-order overtones, the effect of turbulent viscosity will be relatively weak 
(please refer to Fig.~\ref{fig3} in Paper I). Besides, LRV will tend to pulsate at 
fundamental and low-order overtones because $\omega \tau_c \ll 1$ in most 
regions of LRV interiors. Here $\tau_c$ is the convective inertial time scale. 
As the frequency increases, the effects of turbulent viscosity increase 
rapidly [the turbulent viscosity is proportional to 
$\omega \tau_c /(1+\omega^2 \tau_c^2)$]. Thus their high overtones tend to be 
pulsationally stable (prease refer to  Fig.~\ref{fig4} in Paper I). For 
RHBS, since the amplitude growth rates are about 2--3 orders of magnitude less 
than those of LRV, the effects of turbulent viscosity become relatively 
important (see Fig.~\ref{fig4}), and will damp the pulsation. The tremendous 
difference of mass-luminosity ratios is another reason for the distinct 
properties of the two types of low temperature variables.
\item The distributions of $\omega \tau_c$ are totally different in the interiors 
of these two types.
In fact, the interaction between convection and oscillations is very complicated. 
The effects of convection on stellar pulsation depend not only on the variation 
amplitudes of convective variables, but also on their relative phases. 
Both the amplitude and phase depend critically on the ratio of 
convective inertial time scale ($\tau_c$) and that of pulsation, i.e. 
$\omega\tau_c$. Comparing the integrated work plot of the present work
(Fig.~\ref{fig4}) with that of long period variables in our previous paper 
(Paper I), we can see that, the quantity $\omega\tau_c$ is very different in 
the interiors of these two types of stars. For the LRV, $\omega\tau_c<1$ 
in most part of the stellar interior except the very outer part near stellar surface; 
while for the RHBS $\omega\tau_c\gg 1$ in almost the whole stellar interior. Referring 
to Fig~3 of our previous work (Paper I), it is clear that the hydrogen and 
helium first ionization regions are the main excitation region of stellar 
pulsation for the LRV, where convection tends to destabilize pulsation, 
while the turbulent viscosity is the main damping 
mechanism for high-order overtones, due to turbulent viscous dissipation 
($\propto \omega\tau_c/(1+\omega^2_r\tau^2_c)$). When $\omega\tau_c\ll 1$, 
the turbulent viscosity is proportional to $\omega\tau_c$, which increases 
very quickly with increasing pulsational frequency. It becomes opposite 
for the RHBS, for which $\omega\tau_c\gg 1$ holds in most part of the 
stellar interior. The excitation due to the first ionization region of helium 
and a part of ionization region of hydrogen is restrained,
and the damping effect of turbulent viscosity on high-order overtones also 
diminishes. This tends to stabilize the fundamental and the low-order overtones 
of the RHBS while the high-order overtones become pulsationally 
unstable. Following such discussion we conclude that the difference in 
$\omega\tau_c$ is a reason for the distinct properties of the two types of 
low temperature variables. As for the question of why $\omega\tau_c$'s are so 
different in the two type of stars, we understand that it is due to the 
convective inertial time scale $\tau_c\propto H_P/x$.
Here $H_P=r^2P/GM_r\rho\propto r^2 P/M_r\rho$ is the local pressure scale 
height and $x$ is the r.m.s. component of the turbulent velocity. Relative to the 
RHBS, the LRV have higher central condensation. Towards stellar 
center, $r$ goes much faster with increasing temperature. Fig.~\ref{fig6} 
plots $r$, $H_P$ and $\omega\tau_c$ of fundamental mode versus 
temperature for a RHBS model and a LRV model respectively. It 
is clearly shown in  Fig.~\ref{fig7} that there exists very large difference 
between the two types of low temperature variables of different luminosities. 
Following the above discussion, we can easily understand the reason
of the distinct properties of the two types of low temperature variables 
although their variabilities are both affected critically by convection. 
Fig.~\ref{fig5} plots the variations of the integrated work of the first,
second, and fourth overtones versus depth for a $T_e=5150K$ RHBS 
model. From this we see that, unlike luminous red stars, the 
turbulent viscosity component $W_{vis}$ does not increase quickly
with $\omega_r$. 
\end{enumerate}
\section{The Spatial Oscillations of the Convective Variables}\label{sec3}
In the theoretical calculation of the steller oscillations, if the local theory
is used to treat the coupling between the convection and pulsation, the 
troublesome spatial oscillations of thermodynamic variables will occur 
(Keeeley, 1977; Baker and Gough, 1979; Gonczi and Osaki, 1980). These spatial 
oscillations are due to the local treatment of the convection, so we could 
expect that it will automaticly disappear if the non-local treatment of convection 
is adopted. Unfortunately, there still exist intense spatial oscillations of 
convective and thermodynamic variables in our previous work on the solar 
radial p-mode oscillations, although a non-local convection theory was used
(Cheng and Xiong, 1997). Now, we use an asymptotic analysis to 
inspect the cause of spatial oscillations. It is necessary to review 
briefly the main points of our non-local convection theory first. Our 
convection theory is a statistical theory of correlation functions. We 
derived a set of dynamic equations of the auto-  and cross-correlations of 
the turbulent velocity and temperature fluctuations. For example, the 
dynamic equation for the auto-correlations of the turbulent velocity is as 
follows (Xiong, Deng and Cheng, 1997),

\bqa
\lefteqn{\DDt \overline{w'^iw'^j}+\overline{w'^iw'^k}\nabla_{\! k}\bar{u^j}
        +\overline{w'^j w'^k}\nabla_{\! k}\bar{u^i} }
\nonumber \\
 & &\mbox{} -\alpha (\gge \overline{w'^j T'\over{\bT}}
        +\ggf \overline{w'^i T'\over{\bT}}) (\nabla_{\! k}\bPhi+\Dt{\bar{u_k}})
\nonumber \\
 & &\mbox{} +\brho \nabla_{\! k}(\brho \overline{u'^k w'^i w'^j})
        = -4 \sqrt{3}\eta_e x\overline{w'^i w'^j}/l_e,
\label{Eq13}
\eqa
where $x=\sqrt{\overline{w'_i w'^i}/3}$ and $l_e$ is the average size of the 
energy-containing eddies. 
We assume that $l_e \sim c_1 H_p$ ($H_p$ is the local pressure scale height).
The right-hand side term of Eq.~(\ref{Eq13}) is the turbulent dissipation, 
which determines the efficiency of the convective energy transport. Thus, 
$c_1$, which is a convective parameter relevant to the turbulent dissipation, 
functions like the parameter $\alpha =l/H_p$ in the mixing-length theory 
(Bohm-Vitense, Z. 1958). In our theory, the efficiency of the convective energy 
transport is approximately the same as letting $\alpha=l/H_p=2c_1$ in the 
oringinal mixing-length theory.
\par
$\brho \overline{w'^i w'^j}$ is the turbulent Reynold stress. We can decompose 
it into an isotropic component (turbulent pressure) and a nonisotropic 
one (turbulent viscosity). An expression for the turbulent viscosity 
very similar in form to the formula of Stokes viscosity has been derived 
(Xiong, Deng and Cheng, 1997). To reveal the nature of the non-local 
convection, we consider only the isotropic component of Eq.~(\ref{Eq13}), i.e.

\bqa
\lefteqn {{3\over 2}\Dt{x^2} =-x^2 \nabla_{\! k}\bu^k 
        +\alpha V^k(\Dt{\bu_k} +\nabla_{\! k}\Phi)}
\nonumber \\
 & & \mbox{} +\brho \nabla_{\! k}(\brho \overline{u'^k w'_i w'^i})
        -2\sqrt{3}\eta_e x^3/l_e ,
\label{Eq14}
\eqa
where $V^k=\overline{w'^k T'/bT}$ embodies the correlation of the turbulent velocity 
and temperature fluctuations. Eq.~(\ref{Eq14}) is the dynamic equation of 
turbulent pressure. But it can also be considered as the equation of conservation 
of turbulunt kinetic energy. The left-hand side of Eq.~(\ref{Eq14}) is the 
rate of turbulent kinetic energy, which equals the sum of the terms on the 
right-hand side. The first one is 
the rate of regular kinetic energy of averagy motion converted to turbulent 
kinetic energy. The second term is the rate of thermal energy converted to 
turbulent kinetic energy through the work done by buoyancy. The last term is 
the rate of turbulent kinetic energy converted to thermal energy due to the 
molecular viscosity. The triple correlations in the brackets of the third term 
are the turbulent kinetic flux, which represents the non-local turbulent kinetic 
energy transport by convection. These triple correlations represent all the 
effects of non-local convection. They reveal the essential
difference between local and non-local theory: If one ignores all the triple 
correlations, the theory will return to the local convection theory 
(Xiong, 1977). In our earliest works, we used a simplified gradient-type 
diffusion approximation (Xiong,1979), i.e.

\bq
\overline{u'_{\! k}w'_i w'^i}=-lx\nabla_{\! k}\overline{w'_i w'^i}=-3lx \nabla_{\! k} x^2,
\label{Eq15}
\eq
where $l$ is the diffusion length. We assume $l=c_2 \sqrt{3} H_p/4$. $c_2$ is 
another convective parameter relevant to the non-local turbulent diffusion. 
The e-folding length of convective overshooting is about $1.4 \sqrt{c_1 c_2}H_p$. 
Let $c_2=0$, then we get back to local convection theory.
\par
Eq.~(\ref{Eq15}) is introduced only as a physically reasonable assumption. 
Recently, we have derived a set of dynamic equations for triple correlations 
(Xiong, Deng and Cheng, 1997). These equations are very complex: The complete 
set of the radiative fluid dynamic equations are up to the 14th order. In the dynamic 
equations for triple correlations, $\overline{u'_kw'_i w'^i}$ contains not 
only the second correlations of turbulent velocity, but also the cross- 
correlations of turbulent velocity and temperature. For simplicity, omitting 
the non-corresponding second correlations, we get a simplified 
time-dependent equation for triple correlations:

\bq
\Dt{\overline{u'_k w'_i w'^i}} + \overline{u'_k w'_i w'^i} 
        =-xl \nabla_{\! k} \overline{w'_i w'^i}
	  =-3xl \nabla_{\! k} x^2.
\label{Eq16}
\eq
In our calculation of the solar radial p-modes oscillations (Cheng and Xiong, 
1997), a set of formulae similar to Eq.~(\ref{Eq16}) had been used to treat 
all triple correlations ($\overline{u'_{\! k} T'^2/\bT ^2}$ and 
$\overline{u'_{\! k} w'^i T'/\bT}$). But the results are just contrary to 
our expectation: The strong spatial oscillations of the convective and related 
thermal variables existed persistently.
\par
For radial pulsation, Eqs.~(\ref{Eq14}) and (\ref{Eq16}) can be rewritten as 
follows

\bqa
\lefteqn {{3\over 2}\Dt{x^2}-{x^2\over\rho}\Dt{\rho}-
BV\left\{{{GM_r}\over r^2}\left[ 1-{\tau_c\over 4}
\Plnr{\left({u_r\over r}\right)}+\Dt{u_r}\right]\right\} }
\nonumber\\
 & &\mbox{} +3\pMr{L_1}+2\sqrt{3}\eta_e{{GM_r\rho}\over{c_1 r^2P}}x^3=0,
\label{Eq17}
\eqa

\bq
\tau_c\left[\Dt{L_1}-L_1\Dt{\ln(\rho r^2)}\right]+L_1
=-{{4\sqrt{3}\pi^2c_2\rho Pr^6x^3}\over{GM_r}}\pMr{\ln x},
\label{Eq18}
\eq
where $3L_1=2\pi r^2 \brho \overline{u'^1 w'_i w'^i}$ is the turbulent kinetic 
energy flux.
In the region where $\omega_r\tau_c\gg 1$, the convective variables $x$ and 
$V$ change much faster than $r$, $\rho$, $T$ and $P$ of the average fluid fields.
Therefore, under the first approximation for the purpose of studying the 
asymptotic properties of the convective variables, we can ignore the slow 
variations of $r$, $\rho$, $P$ and $T$ of the fluid fields, and simply 
take them as constants. Bearing 
this in mind and linearizing Eqs.~(\ref{Eq17}) and (\ref{Eq18}), we have 
the following approximated and linearized equations of pulsation,

\bqa
\lefteqn{
{{\sqrt{3}c_2Px^3}\over{4GM_r\rho}}\ddlnr{\ln x}\dDlnr{\left({{\delta L_1}\over L_1}\right)}
+{1\over{\rho r^3}}{{\delta L_1}\over L_1}\dDlnr{\left({{\sqrt{3}c_2r^3Px^3}\over{4GM_r}}\ddlnr{\ln x}
\right)} }\nonumber\\
 & &\mbox{} -{{GM_r}\over r^2}\left[ \sqrt{3}\eta_e\left( 2+i\omega\tau_c\right)
            {{\rho x^3}\over{c_1 P}}{{\delta x}\over x}-{1\over 3}BV{{\delta V}\over V}\right]=0,
\label{Eq19}
\eqa

\bq
{{\sqrt{3}\pi c_2r^3Px^3}\over{GM_r}}\left\{\dDlnr{\left({\delta x}\over x}\right)
 + \ddlnr{\ln x}\left[ 3{{\delta x}\over x}-\left( 1+i\omega\tau_c\right){{\delta L_1}\over L_1}
\right]\right\} =0.
\label{Eq20}
\eq
Besides the convective variables $\delta x/x$ and $\delta L_1/L_1$, Eq.~(\ref{Eq19}) 
still contains another convective variable $\delta V/V$. Its linearized equation 
will have other variables, namely $\delta x/x$ and $\delta Z/Z$, 
where $Z=\overline{T'^2}/\bar{T}^2$ represents the auto-correlation of the 
turbulent temperature fluctuation. In this case, the three auto- and 
self-correlations of the fluctuations of turbulent velocity and temperature are 
closely coupled. We have to solve the 6 linear differential equations for 
the asymptotic solutions, and this is an extremely difficult procedure. Without 
losing any generality, we will omit the terms containing the non-diagonal 
element $\delta V/V$ in Eq.~(\ref{Eq19}) when we discuss Eqs.~(\ref{Eq20}) and (\ref{Eq19}). Replacing 
the variable with $\eta=\ln P$, the two equations can be simplified as

\bqa
\deta{Y_1}+\alpha\left[ 3Y_1-\left( 1+i\omega\tau_c\right)Y_2\right]=0,
\label{Eq21}
\eqa

\bqa
\deta{Y_2}+\left( 3\alpha + \deta{\ln\rho}+2\deta{\ln r}\right) Y_2
 -{{4\eta_e}\over{c_1c_2\alpha}}\left( 2+i\omega\tau_c\right) Y_1=0,
\label{Eq22}
\eqa
where

\bq
\alpha={{{\rm d}\ln x}\over{{\rm d}\ln P}},
~~~Y_1={{\delta x}\over x},~~~Y_2={{\delta L_1}\over L_1}.
\label{Eq23}
\eq
Neglecting the variation of $\rho r^2$ in space and making the 
variable conversion

\bq
Y_2=\sqrt{{{4\eta_e}\over{c_1c_2}}{{2+i\omega \tau_c}\over{1+i\omega \tau_c}}}
{1\over\alpha}X={\beta\over\alpha}X,
\label{Eq24}
\eq
Eqs.~(\ref{Eq21}) and (\ref{Eq22}) can be written as

\bq
\deta{Y_1}+3\alpha Y_1-\beta\sqrt{\left( 1+i\omega\tau_c\right)
\left( 2+i\omega\tau_c\right)}X=0,
\label{Eq25}
\eq

\bq
\deta{X}+3\alpha X-\beta\sqrt{\left( 1+i\omega\tau_c\right)
\left( 2+i\omega\tau_c\right)}Y_1=0.
\label{Eq26}
\eq
The characteristic values for Eqs.~(\ref{Eq22}) and (\ref{Eq23}) are

\bq
\lambda_\pm=-3\alpha\pm \beta\sqrt{\left( 1+i\omega\tau_c\right)
\left( 2+i\omega\tau_c\right)}\approx -3\alpha\pm
\left({3\over2}+i\omega\tau_c\right)\beta,
\label{Eq27}
\eq
and their general solutions are

\bq
Y_1=a_1P^{-3\alpha+{3\over 2}\beta}\cos\left(\beta\omega\tau_c\eta 
+\phi_1\right) +a_2P^{-3\alpha-{3\over 2}\beta}
\cos\left(\beta\omega\tau_c\eta +\phi_2\right),
\label{Eq28}
\eq

\bq
Y_2={\beta\over\alpha}\left[a_1P^{-3\alpha+{3\over 2}\beta}
\cos\left(\beta\omega\tau_c\eta +\phi_1\right)
-a_2P^{-3\alpha-{3\over 2}\beta}\cos\left(\beta\omega\tau_c\eta 
+\phi_2\right)\right],
\label{Eq29}
\eq
where $a_1$, $a_2$, $\phi_1$ and $\phi_2$ are constants. Apparently, the 
solutions Eqs.~(\ref{Eq28}) and (\ref{Eq29}) are of the oscillating type, 
and the wavelength of the spatial oscillation is

\bq
{{2\pi}\over{\beta\omega\tau_c}}H_P\approx 4.68{\sqrt{c_1c_2}
\over{\omega\tau_c}}.
\label{Eq30}
\eq
Within the convective overshooting zone, it follows from the asymptotic 
properties of convective overshooting (Xiong, 1989) that

\bq
\alpha\approx \pm\sqrt{{{1+2\sqrt{3}\eta_e}\over{3\sqrt{3}c_1c_2}}}
=\pm{{0.70175}\over{\sqrt{c_1c_2}}}.
\label{Eq31}
\eq
In the upper overshooting zone of a convective region, 
$\alpha$ is the positive solution. To avoid a too fast outwards growing  amplitude, 
we choose $\lambda=\lambda_+$, i.e. we take $a_2=0$ in Eqs.~(\ref{Eq28}) 
and (\ref{Eq29}). On the other hand, in the lower overshooting zone of a convective 
region, where $\alpha<0$, we take $\lambda=\lambda_-$, i.e. $a_1=0$. 
\par
According to Eqs.~(\ref{Eq24}), (\ref{Eq28}), (\ref{Eq29}) and (\ref{Eq31}), the 
convective variables tend to have an oscillating solution with a growing amplitude 
towards convective overshooting zone, the e-folding length of which is

\bq
\left| -3\alpha\pm{3\over 2}\beta\right|={{0.09278}\over\sqrt{c_1c_2}}H_P.
\label{Eq32}
\eq
\par
There exist the similar spatial oscillating behavior as above for the 
auto-correlation of temperature fluctuation $Z$ and the cross-correlation 
$V$ of velocity and temperature fluctuations. The procedure to follow is also 
the same as described above, so we will not go into theirs details.
\par
Now, it is not difficult to understand why there exist intense spatial 
oscillations in the region where $\omega_r\tau_c\gg 1$. Let's look back at 
Eq.~(\ref{Eq16}). The triple correlations $\overline{u'_{\alpha}w'_i w'^i}$ 
must approach to zero when $\omega_r\tau_c\gg 1$. That is to say, the 
nonlocal convection theory approaches to the local one. Thus the spatial 
oscillations occurring in the calculation of local convection theory will 
certainly come out. To avoid spatial oscillations, we go back to the simple 
gradient type diffusion approximation respectively by Eq.~(\ref{Eq15}). 
The corresponding linear pulsational equation becomes

\bq
\deta{Y_1}+3\alpha Y_1-\alpha Y_2=0.
\label{Eq33}
\eq
Making a new variable conversion

\bq
Y_2=\sqrt{{{4\eta_e}\over{c_1c_2}}\left( 2+i\omega \tau_c\right) }
{1\over\alpha}X={\beta_1\over\alpha}X,
\label{Eq34}
\eq
Eqs.~(\ref{Eq33}) and (\ref{Eq22}) can be written as

\bq
\deta{Y_1}+3\alpha Y_1-\beta_1\sqrt{ 2+i\omega\tau_c}X=0,
\label{Eq35}
\eq

\bq
\deta{X}+3\alpha X-\beta_1\sqrt{ 2+i\omega\tau_c}Y_1=0.
\label{Eq36}
\eq
Their characteristic values are

\bq
\lambda'_\pm=3\alpha\pm\sqrt{2+i\omega\tau_c}\beta_1=3\alpha\pm
\sqrt{2}\beta_1 \left( \cos{\phi}+i\sin{\phi} \right),
\label{Eq37}
\eq
where $2\phi =\arctan{{\omega \tau_c}\over{2}}$. It is clear that if 
$\omega \tau_c \rightarrow \infty$, we have $\phi \rightarrow \pi /4$. 
Therefore, in the upper overshooting zone where $\alpha>0$, we chose the 
solution with $\lambda=\lambda'_+$; while in the bottom overshooting zone where 
$\alpha<0$ we chose the solution with $\lambda=\lambda'_-$. Then the spatial 
oscllations become solutions attenuating rapidly towards the overshooting zone. 
By doing so, we restrained effectively the spatial oscillations in our 
calculations. The trace of spatial oscillations of eigen-functions can still 
be found in the outmost zones where $\omega \tau_c \gg 1$ in Fig.~\ref{fig1}, 
and especially for convection variables ($y_5$--$y_{10}$). The eigen-function 
of turbulent velocity in the outmost stellar zones is plotted on 
Fig.~\ref{fig1}f, where the spatial oscillations are very clear. 
\par
These asymptotic solutions were used as the convective boundary conditions at 
the top and bottom in the numerical calculations of non-adiabatic oscillations.

\section{Discussion and Conclusions}\label{sec4}
We have carried out theoretical calculations of linear 
non-adiabatic oscillations for six series of HB stellar models. The main results 
can be summarized in the following, 
\begin{enumerate}
\item When the coupling between convection and oscillations is ignored, it is 
not possible to interpret the red edge of the RR Lyrae instability strip. All 
the HB models with $T_e\leq 6200K$ are pulsationally unstable from the 
fundamental up to (at least) the 9th overtone;
\item The intrinsic reason for the existence of the red edge of the RR instability 
strip is the coupling between convection and oscillations. Our numerical results 
show that, when taking the coupling into consideration, all the low temperature HB 
models with $T_e\leq 5940K$ have pulsationally stable fundamental and low-order 
overtones ($n\leq 3$). The turbulent viscosity is an important 
damping mechanism of oscillations.
\item For the low temperature red HB star models some of the high-order overtones 
($n\geq 4$) become pulsationally unstable. The pulsational periods of these 
high-order overtones are of the order of a few tenths of a day. This might be 
the clue for the red variables outside the RR Lyrae instability strip 
in globular clusters found by Yao and his collabolators(Yao, 1981, 1986, 1987; 
Yao et al. 1993a,b, 1994).
\item The pulsational properties of RHBS are very different from those of LRV.
In the current works, all low temperature red models with $T_e\leq 5940K$ have 
pulsationally unstable fundamental and the low-order overtones ($n\leq 3$), 
while some of their high-order overtones ($n\geq 4$) are pulsationally unstable. 
Opposite to this, the LRV outside the Cepheid instability strip 
(i.e. the long period variables) are pulsationally stable for all the high-order 
overtones ($n\geq 4$), while the fundamental and the first overtone are possibly 
unstable. The tremendous difference of mass-luminosity ratios is the reason for 
the distinct properties of the two types of low temperature variables.
\item The spatial oscillations of the convective and the thermal variables in 
the local time-dependent convection theory are effectively restrained in the 
present paper, where a nonlocal time-dependent convection theory is used to 
treat the coupling between convection and oscillations.
\end{enumerate}
\par
Comparing Tables 2 and 3 we have found that, the effects of convection on the instability 
of such stars still cannot be ignored for the blue HB stars ($T_e\geq 7000K$), 
although the convective envelope is already very shallow and the convective
energy transport becomes negligible ($L_c/L\ll 1$). These effects are mainly
due to the dynamic coupling between convection and oscillations, other than the
thermodynamic coupling, because in the outermost layer of the blue HB stars,
a certain amount of turbulent pressure remains, even if the convective energy 
transport is negligible there. We made a numerical test, in which the dynamic 
coupling was ignored. The results show that all the low-order overtones become 
pulsationally stable. We are reluctant to draw any further conclusion on the 
pulsational properties of the blue HB stars because there are some undetermined 
factors, i.e.
\begin{enumerate}
\item There are spatial oscillations in the stellar atmosphere (Fig.~\ref{fig1}f).
We cannot estimate quantitatively their influence on the results of numerical 
calculations. Their effect, we think, may be unimportant to the cool RHBS, because 
they have extended convective envelopes. There are no any spatial oscillations 
in the main part of the convective zone except the atmosphere 
(Fig.~\ref{fig1}). However, convection appears only in a limited surface region 
and the pulsational stability coefficient is small for the blue HB stars. The 
spatial oscillations become, maybe, more important in comparison with the RHBS.
\item The undetermined structure of the turbulent atmosphere of stars. The 
atmosphere has an unnegligible influence on the pulsational stability of stars, 
especially for the blue HB stars, which have small pulsational stability. It 
is very difficult to construct an exact enough model of turbulent atmosphere.
\end{enumerate}
\par
This problem deserves future investigation. The high-precision photometry 
of variables is available for us to determine whether there are low-order 
unstable modes in the blue HB stars.
\acknowledgments
The present work is supported in part by the contract 19573018 of the 
National Natural Science Foundation of China(NSFC).

\newpage
Figure Captions
\figcaption[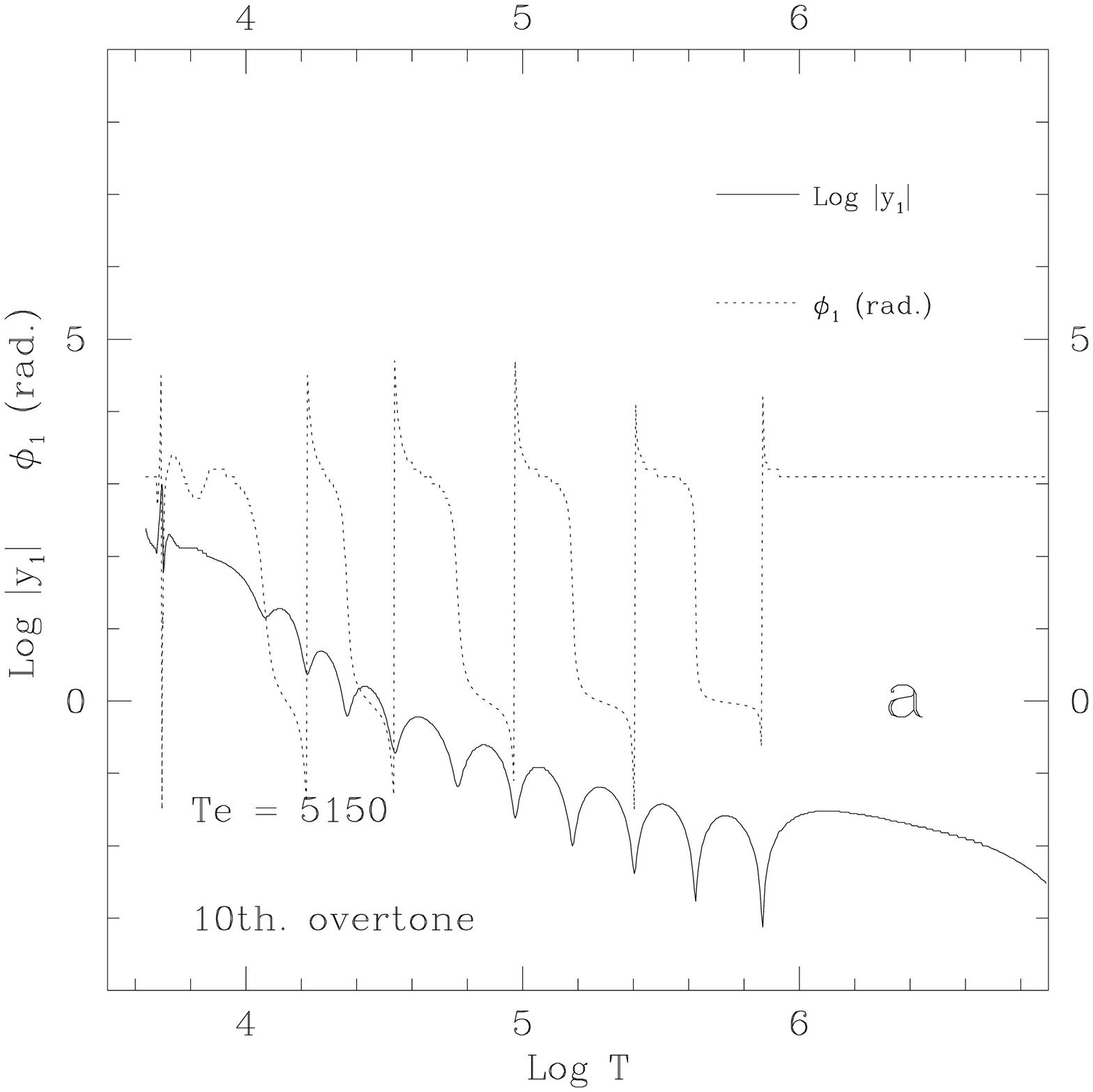,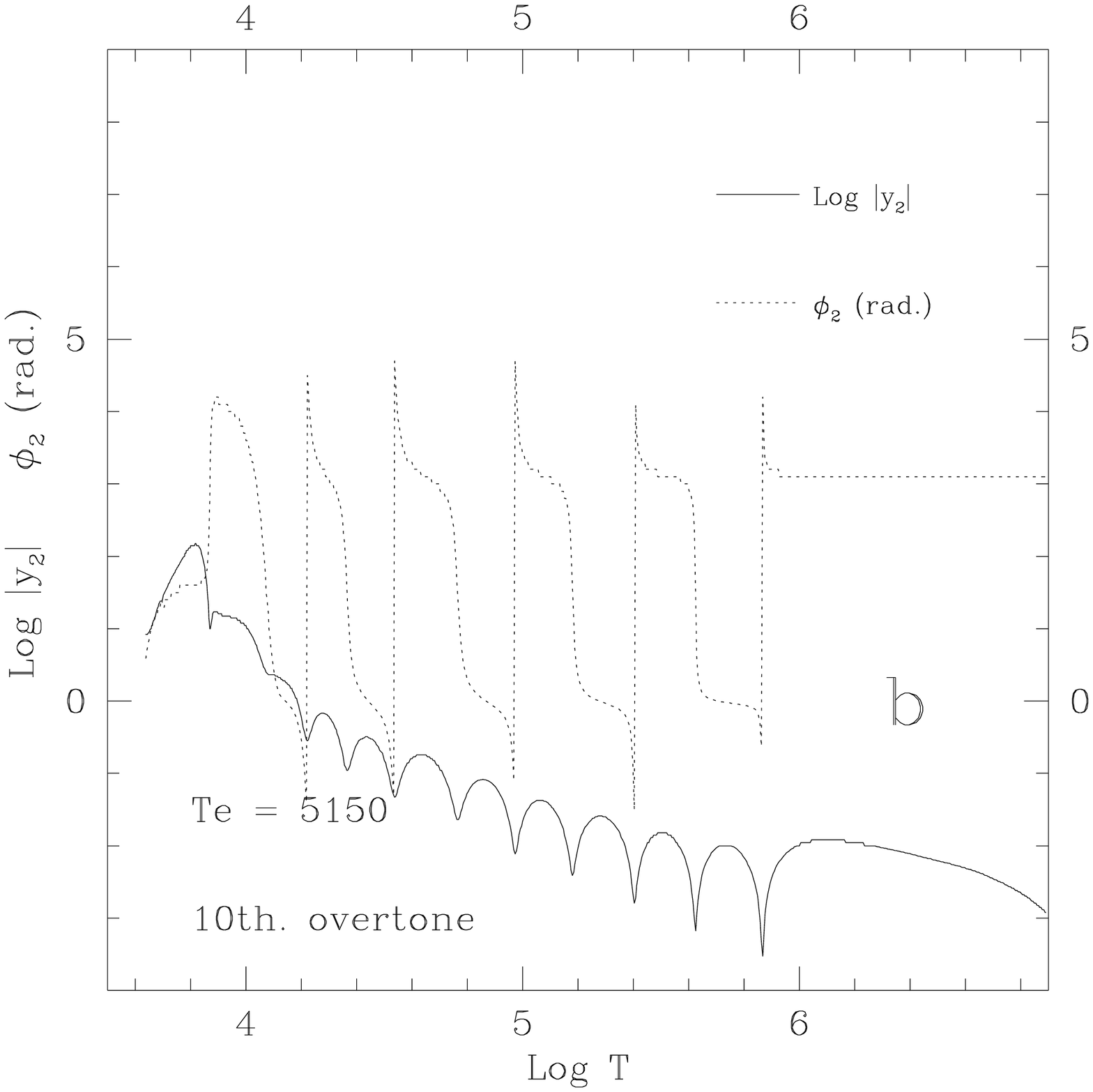,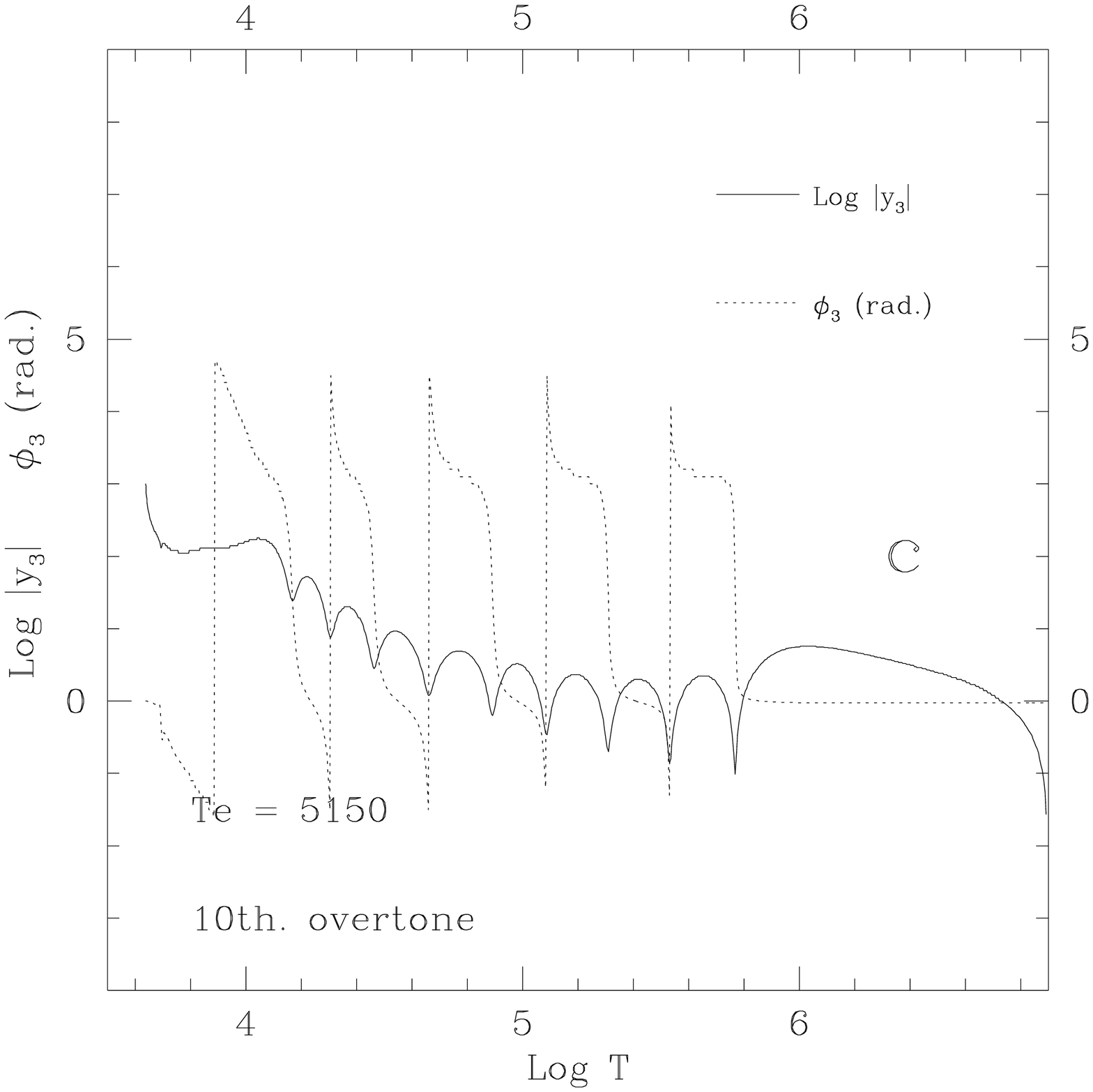,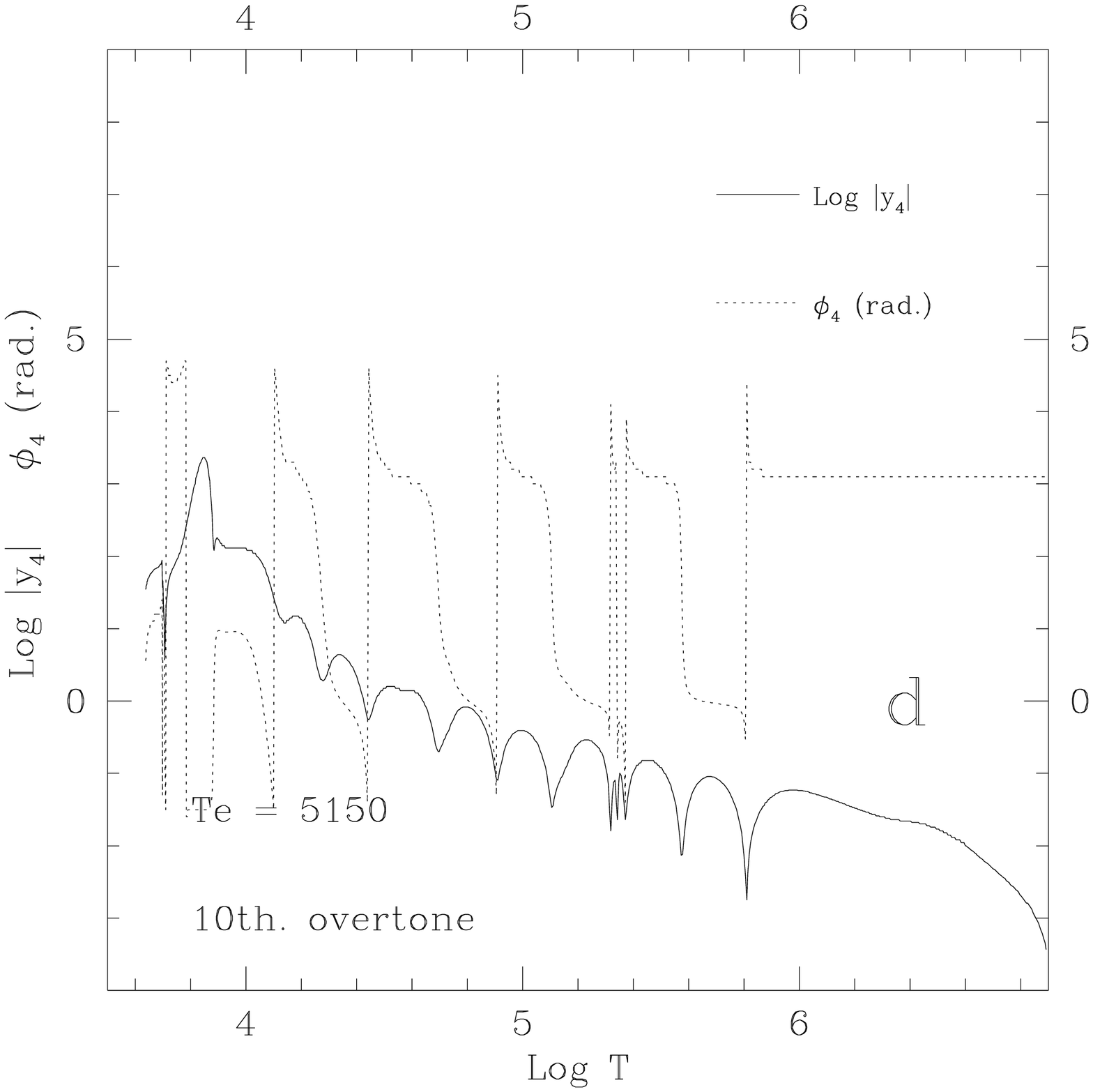,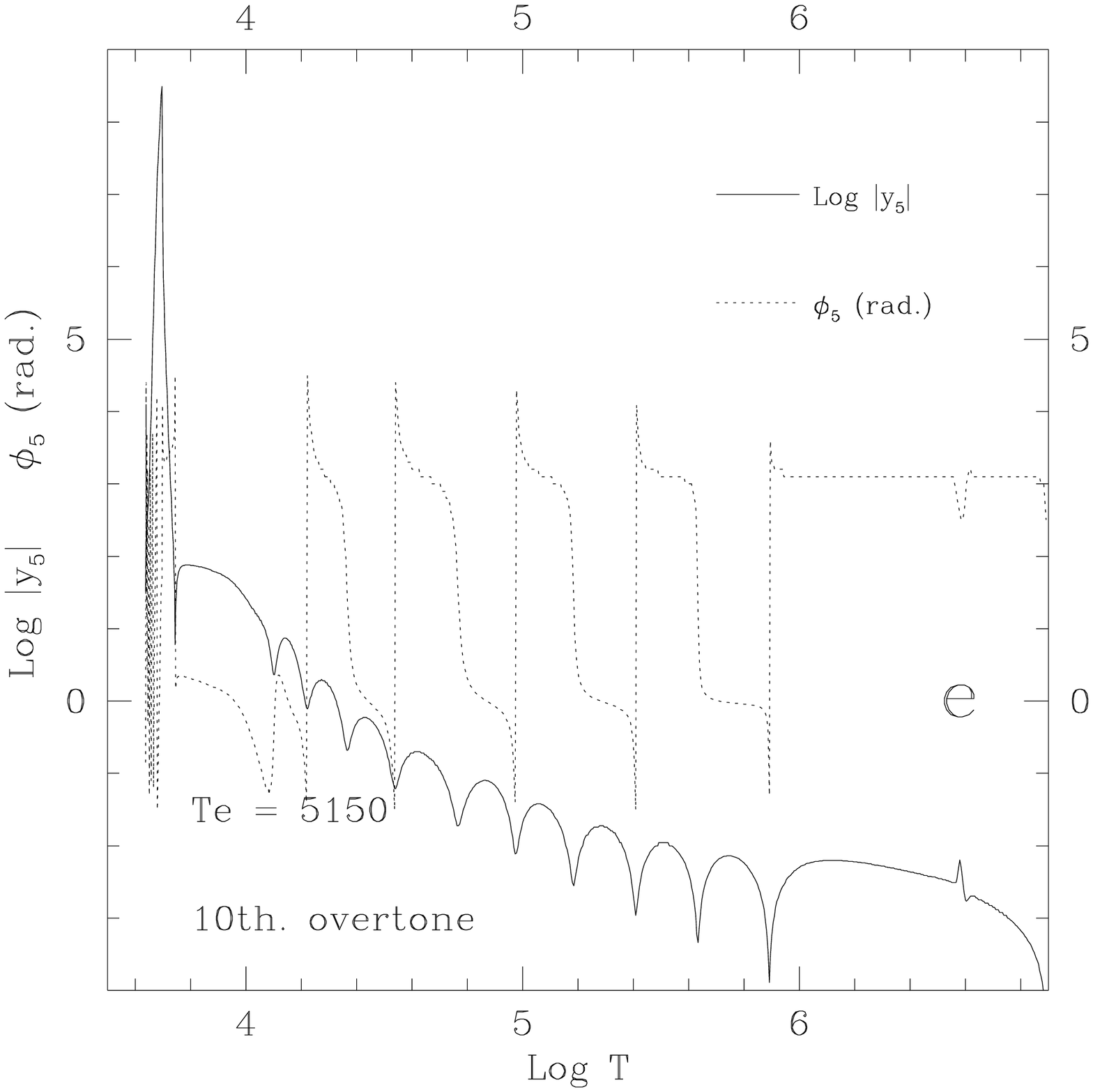,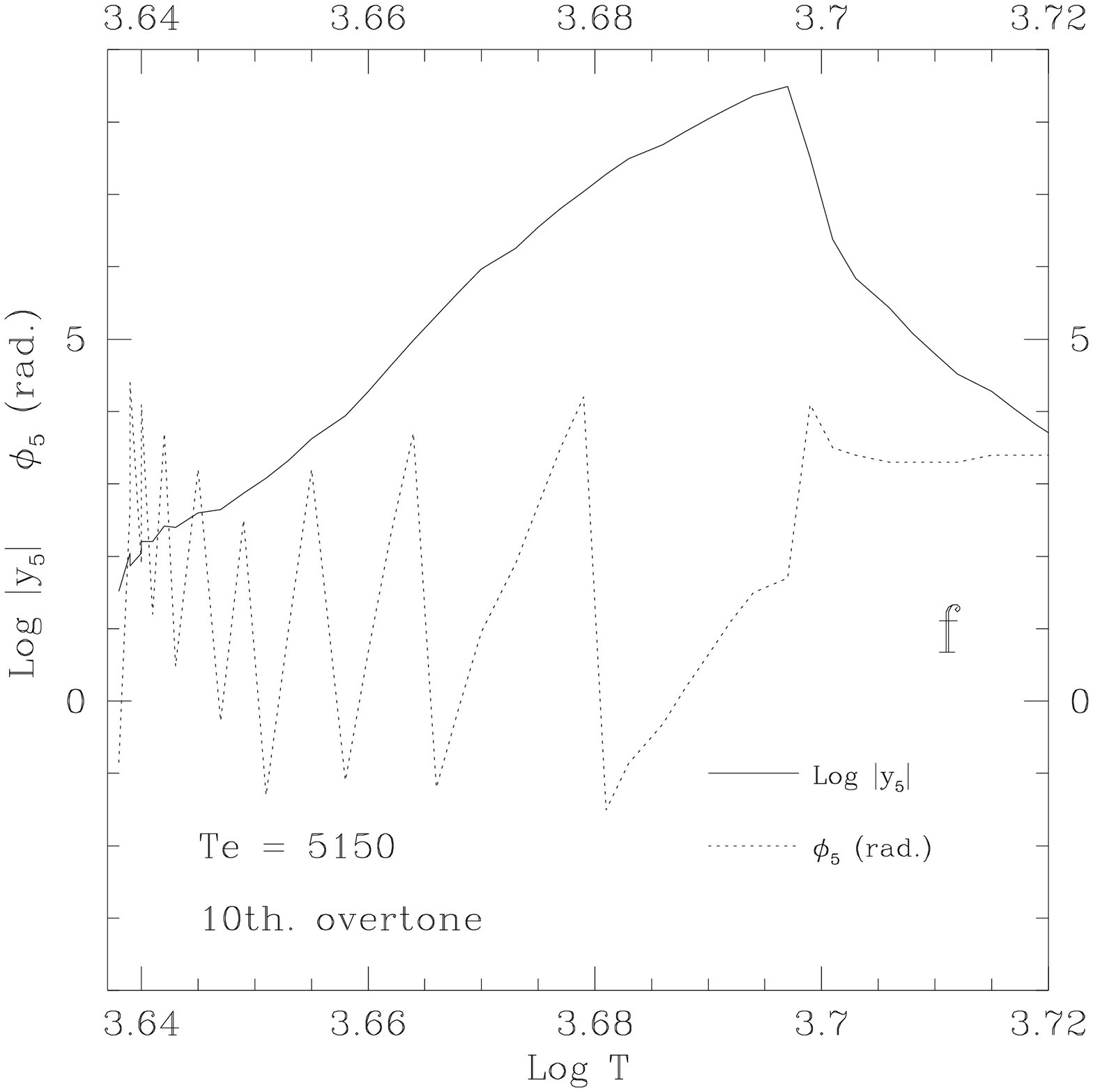]{
The normalized eigen-functions of the 10th overtone for the coldest HB star
model in series 2.
\label{fig1}}

\figcaption[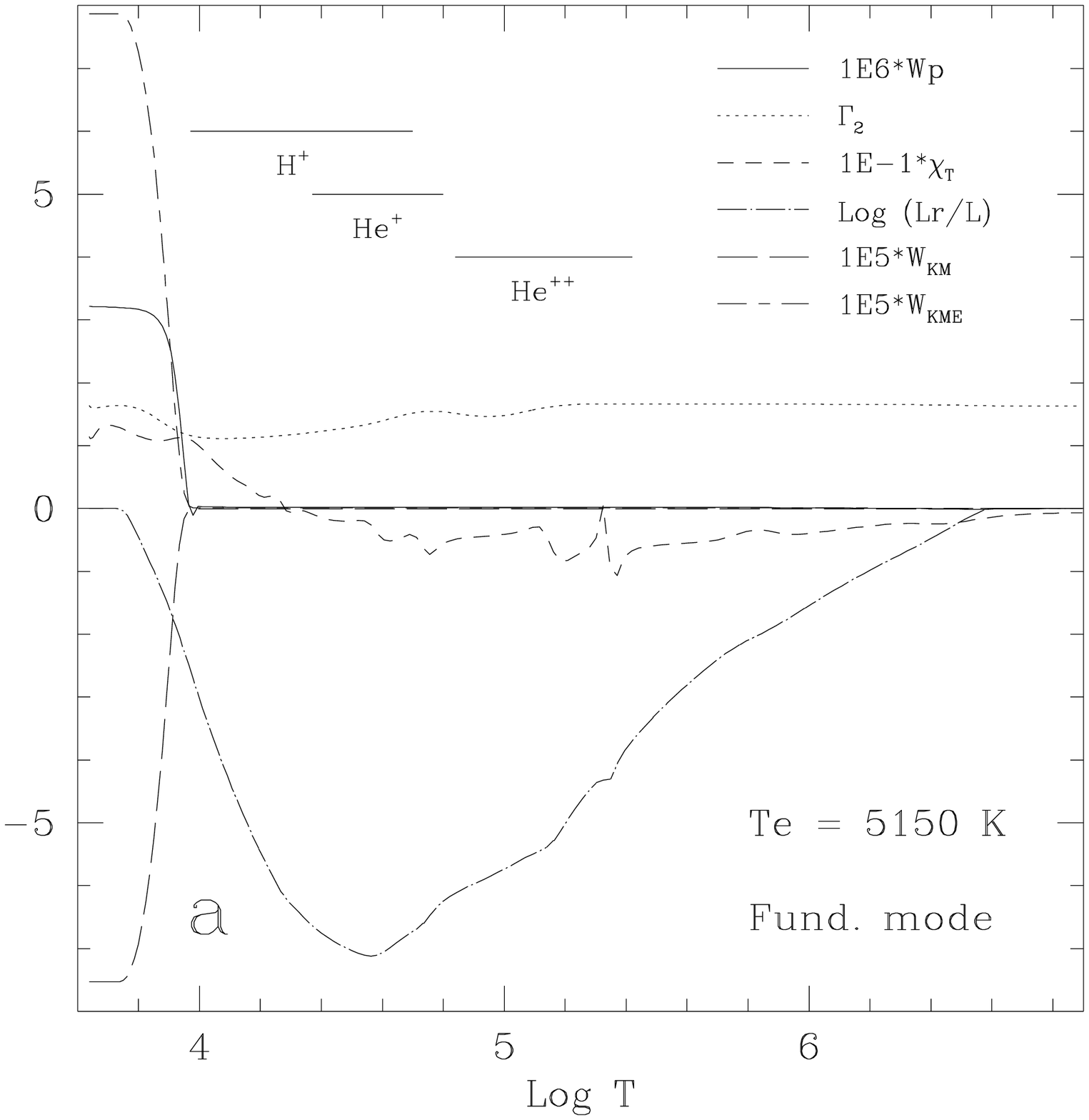,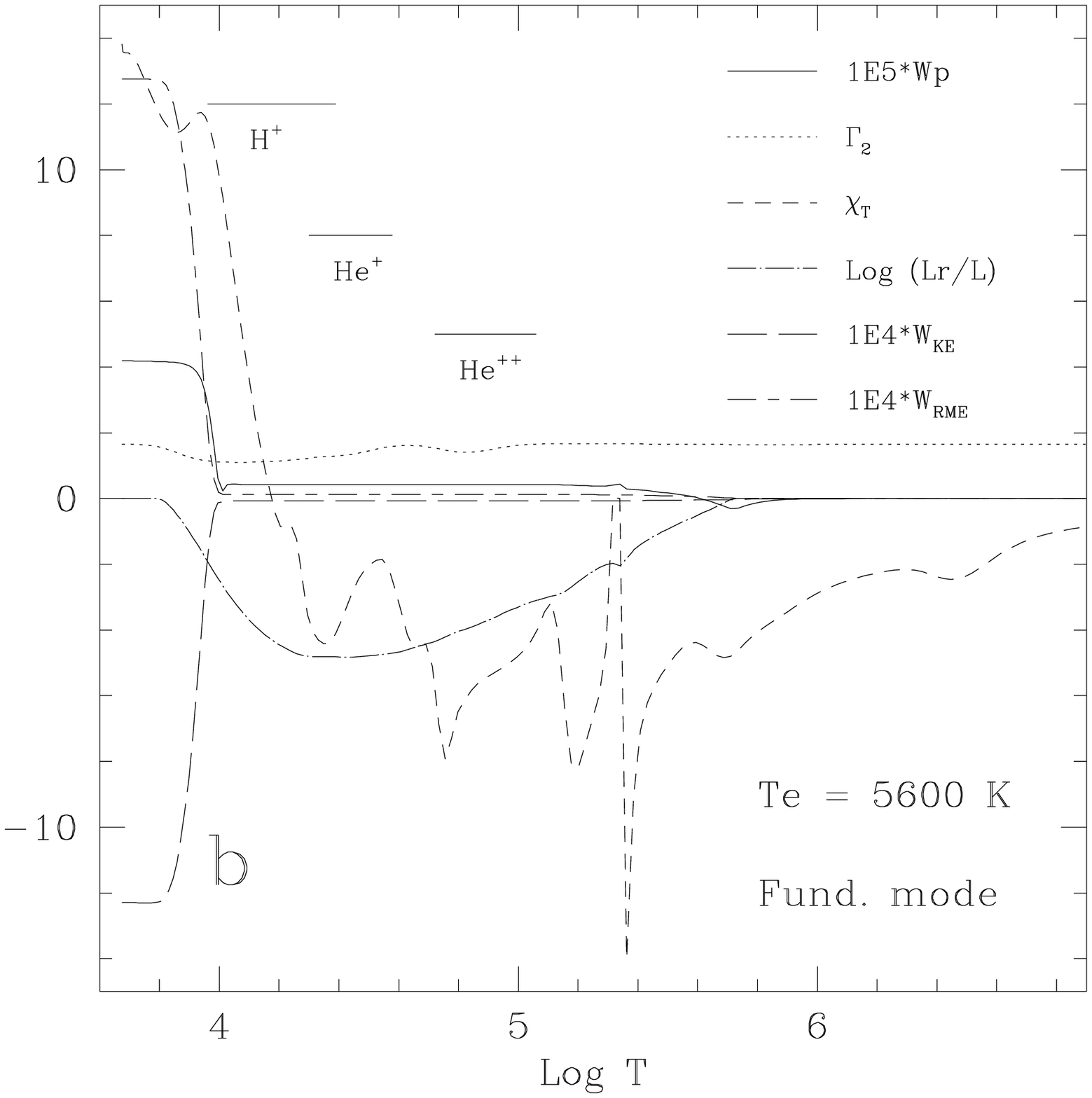,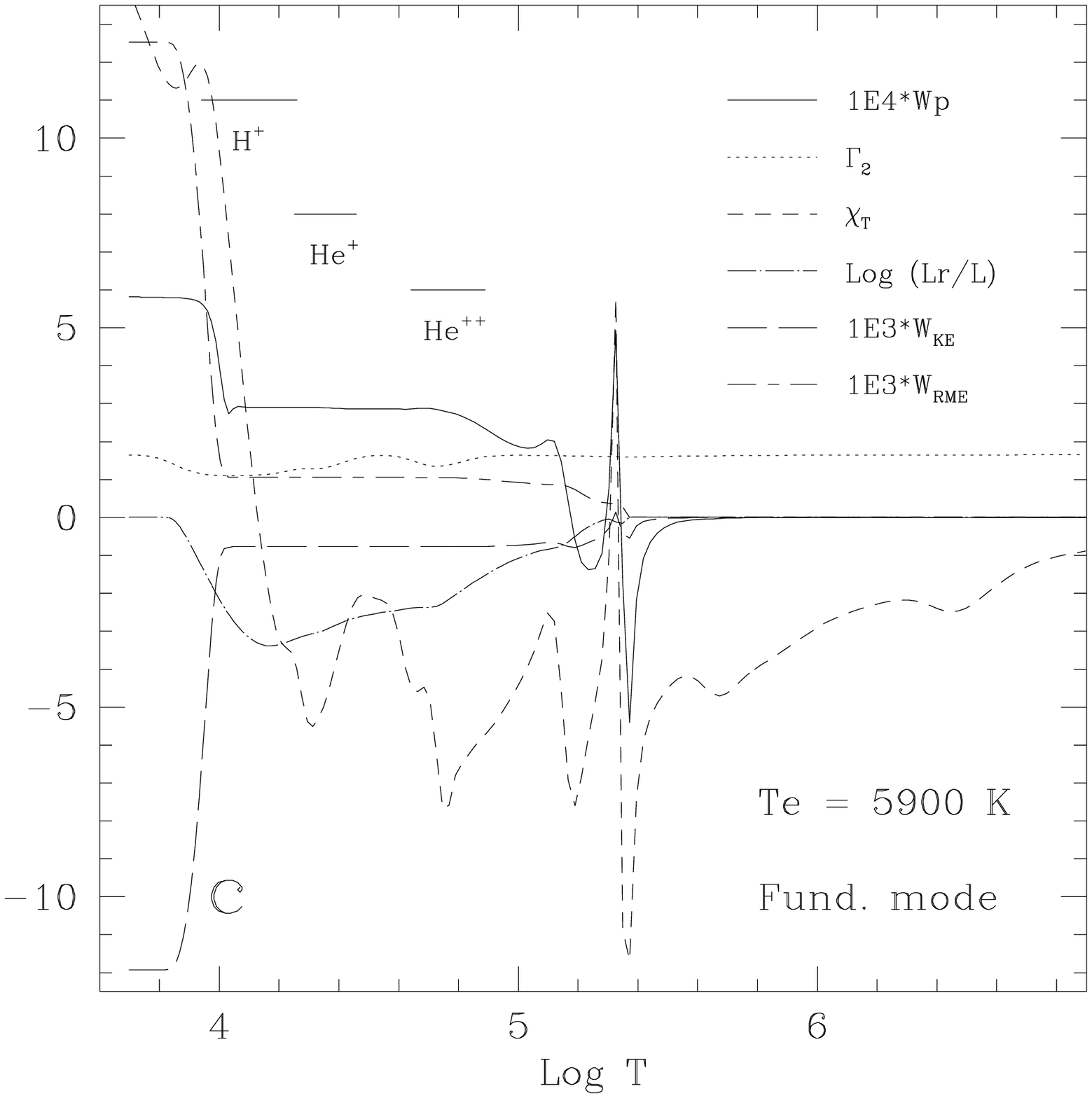,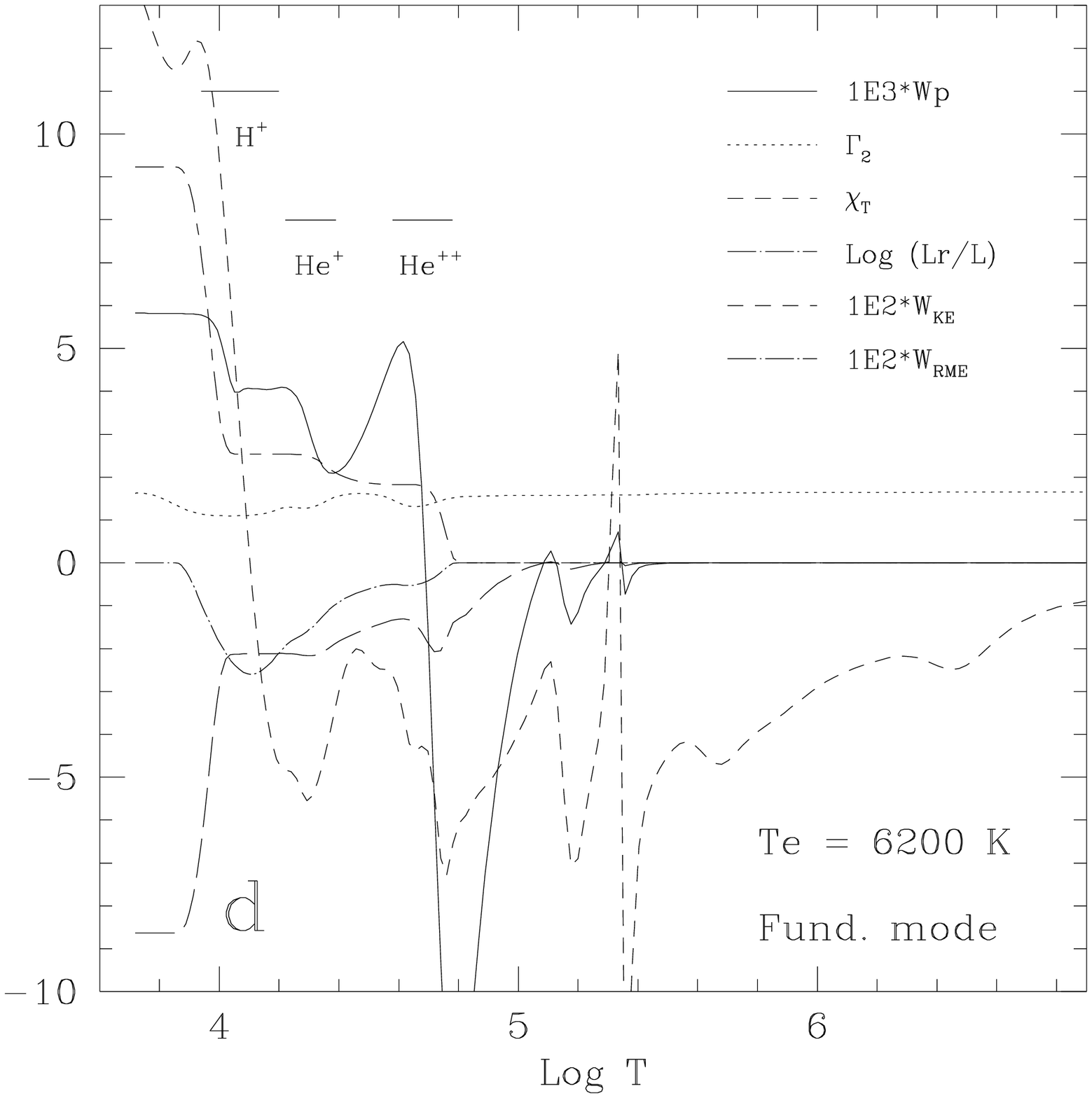,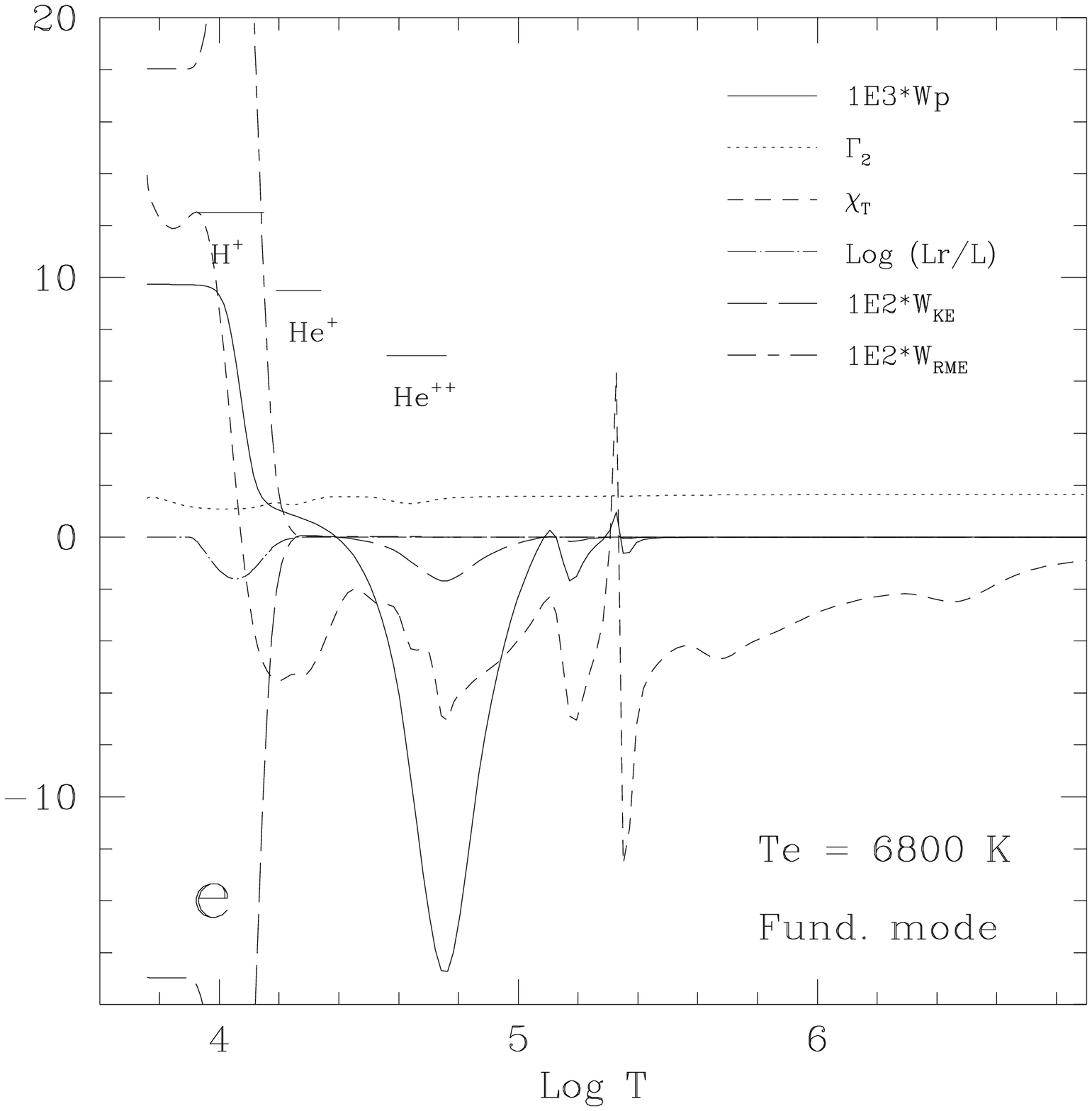,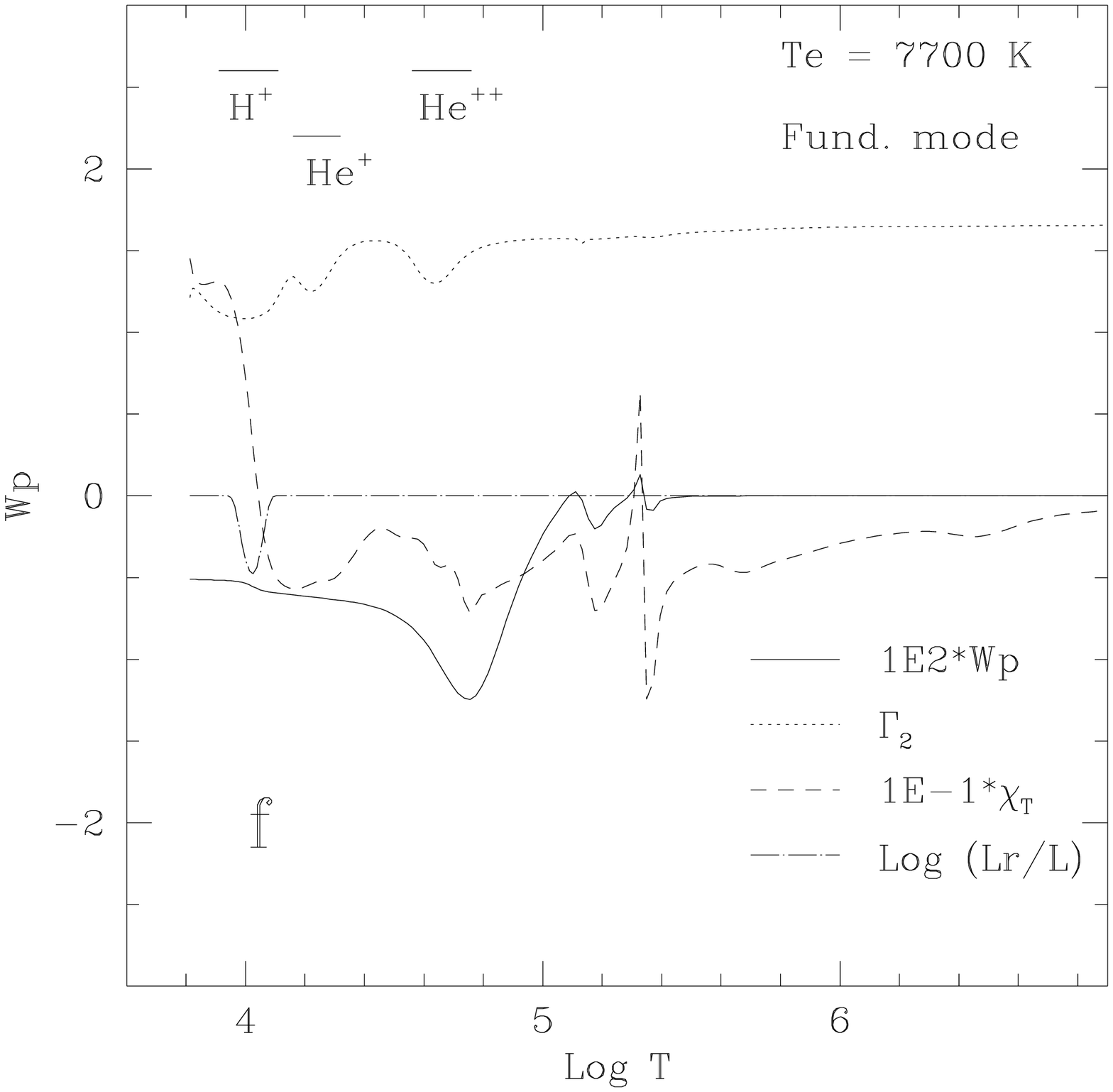]{
The integrated work $W_P$ versus depth for six different HB star models. 
The coupling between convection and oscillations is ignoreed. $W_{KM}$ and 
$W_{RME}$ are respectively the contribution components of the $\kappa$ mechanism
(KM) and radiative modulation excitation (RME).The radiative energy 
flux $L_r/L$, the adiabatic index $\Gamma_2$, and the partial differential 
of opacity with respect to temperature, $\chi_{_T}$ are also plotted. Three 
horizontal lines indicate the locations of the ionization regions of H and He.
\label{fig2}}

\figcaption[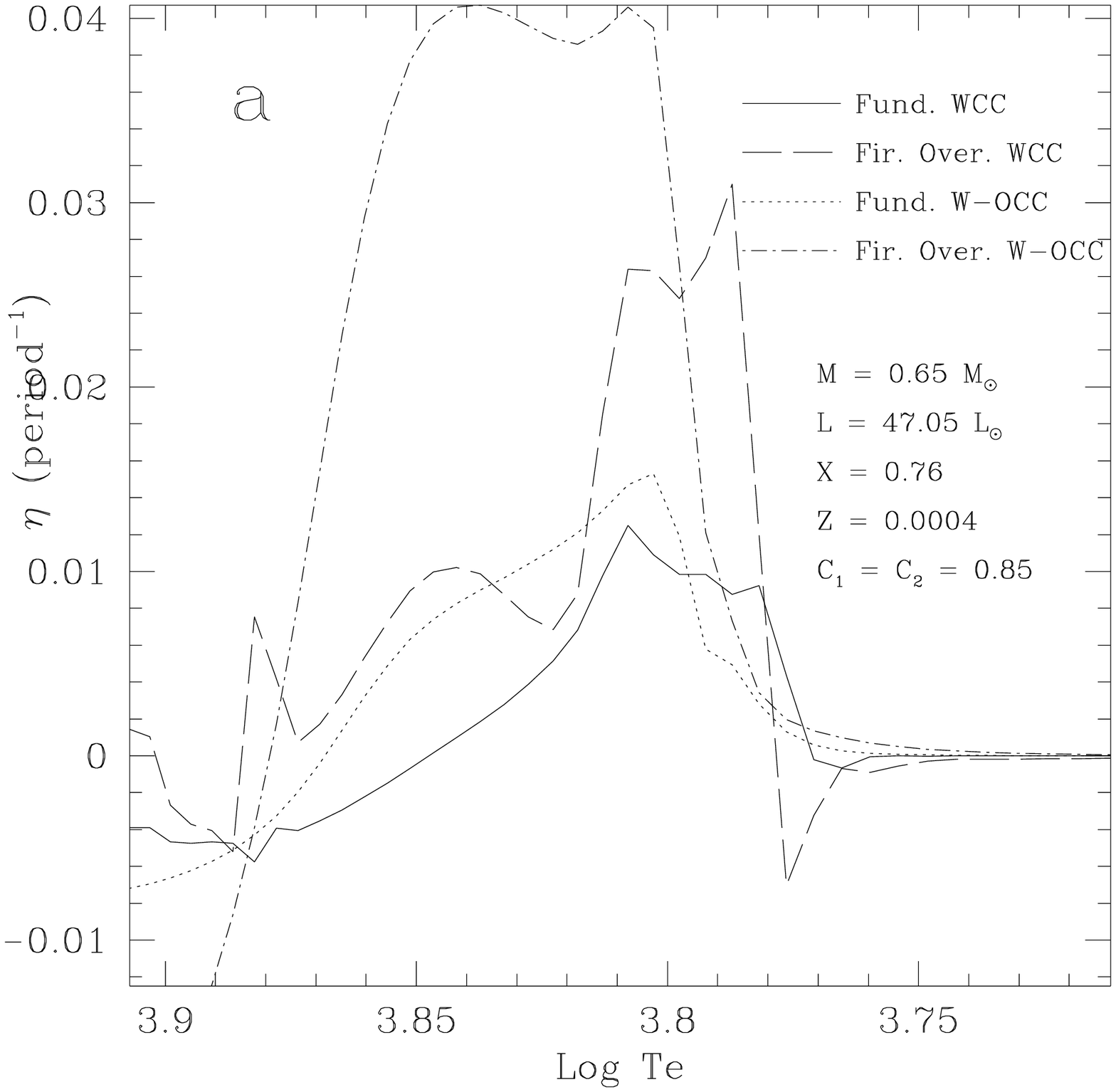,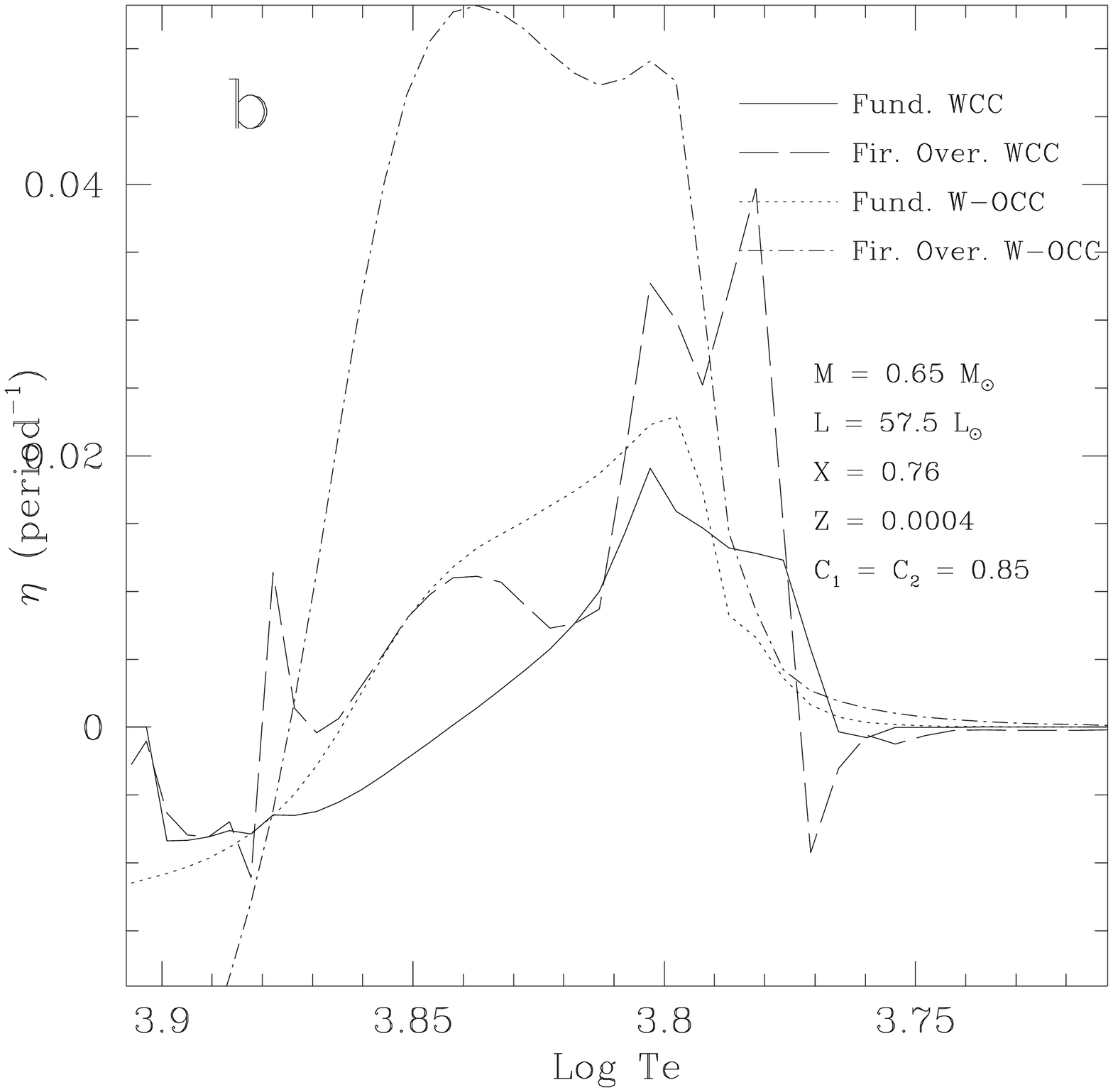,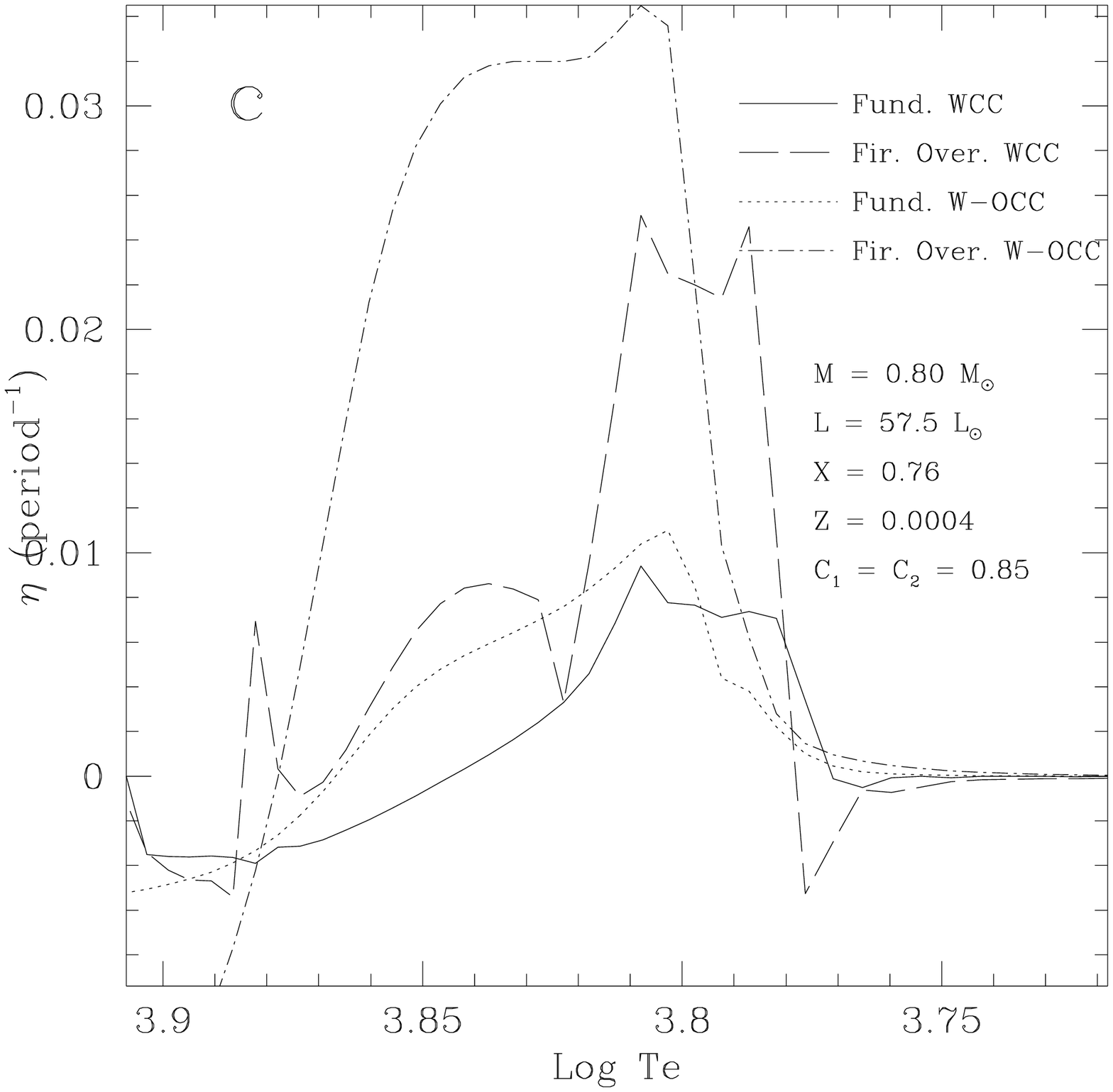]{
The variation of the growth rates of the fundamental and the
first overtone versus effective temperature for three model series of HB stars.
The solid and long-dashed lines show the cases with consideration of the coupling
between convection and oscillations (WCC), while the dotted and the dot-dashed
lines shows the cases ignoring the convection coupling (W - OCC).
\label{fig3}}

\figcaption[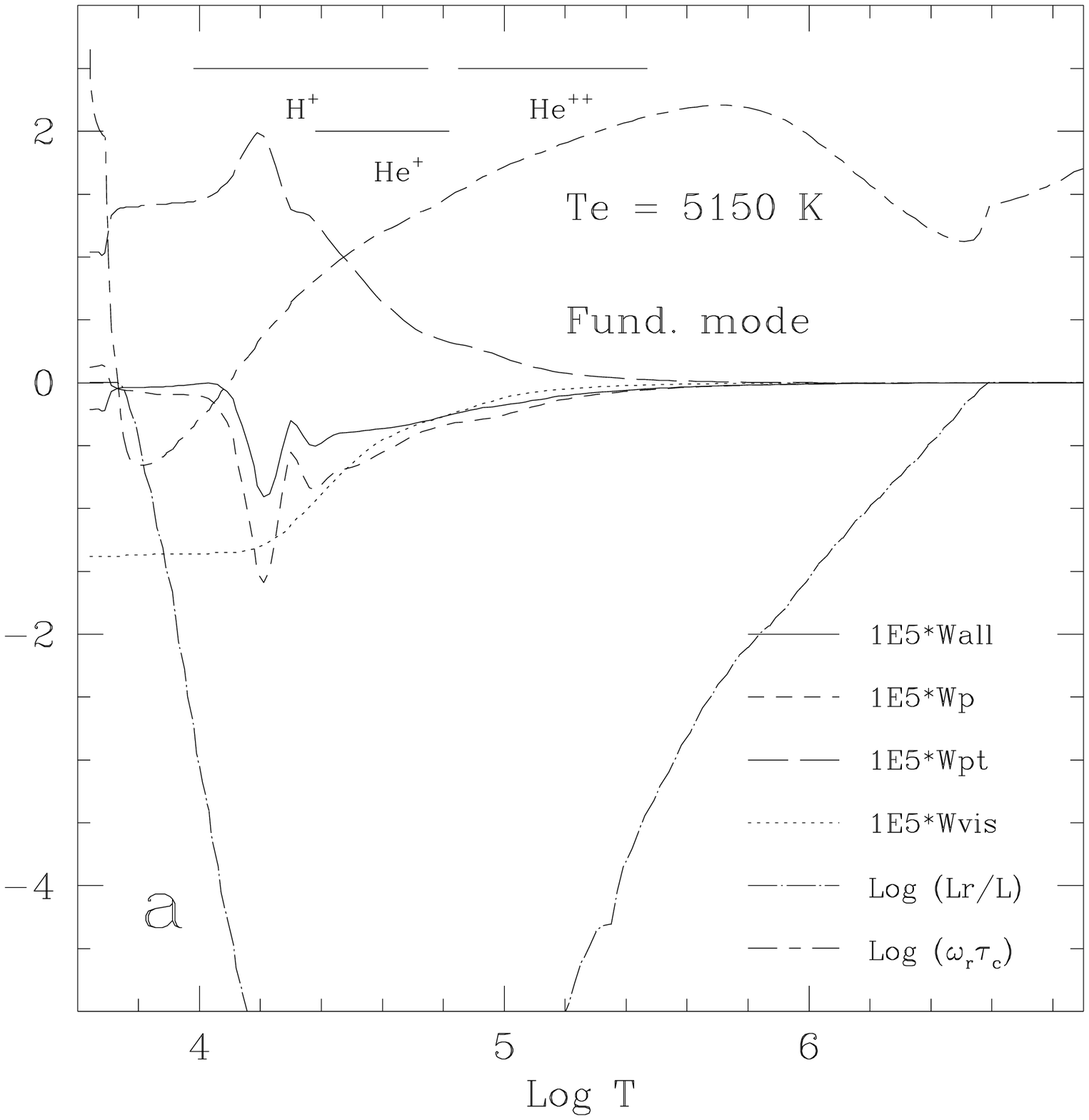,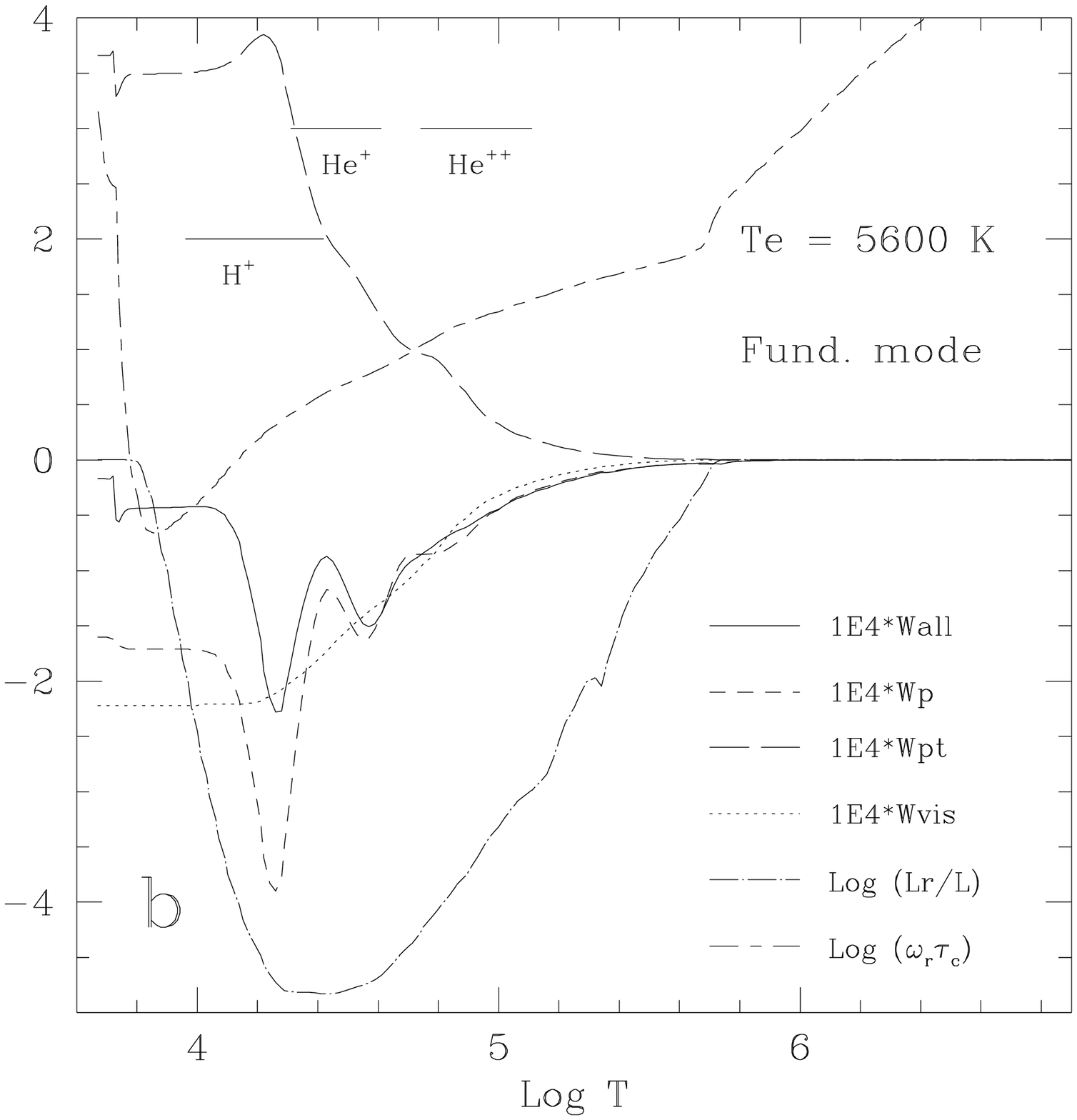,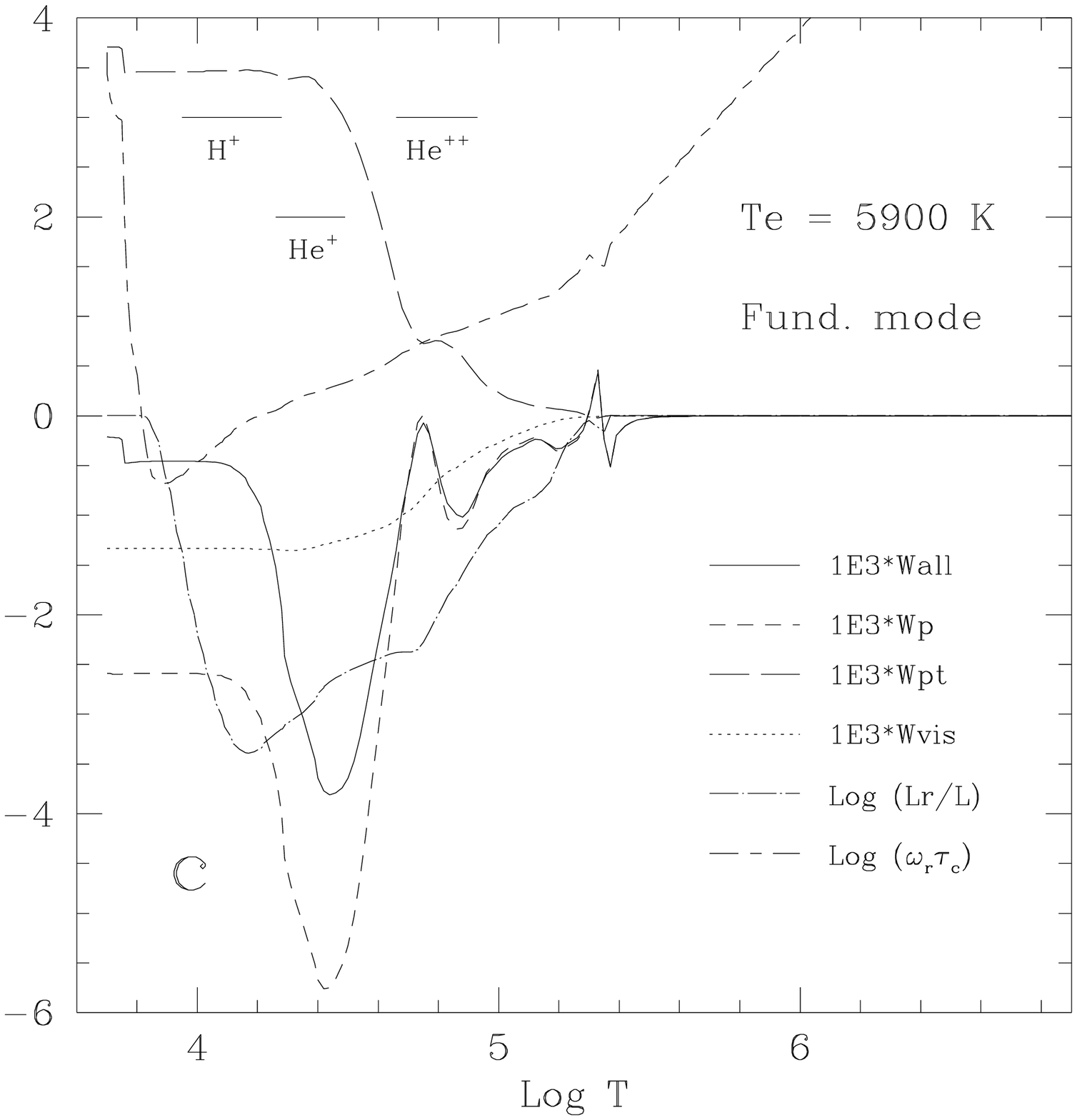,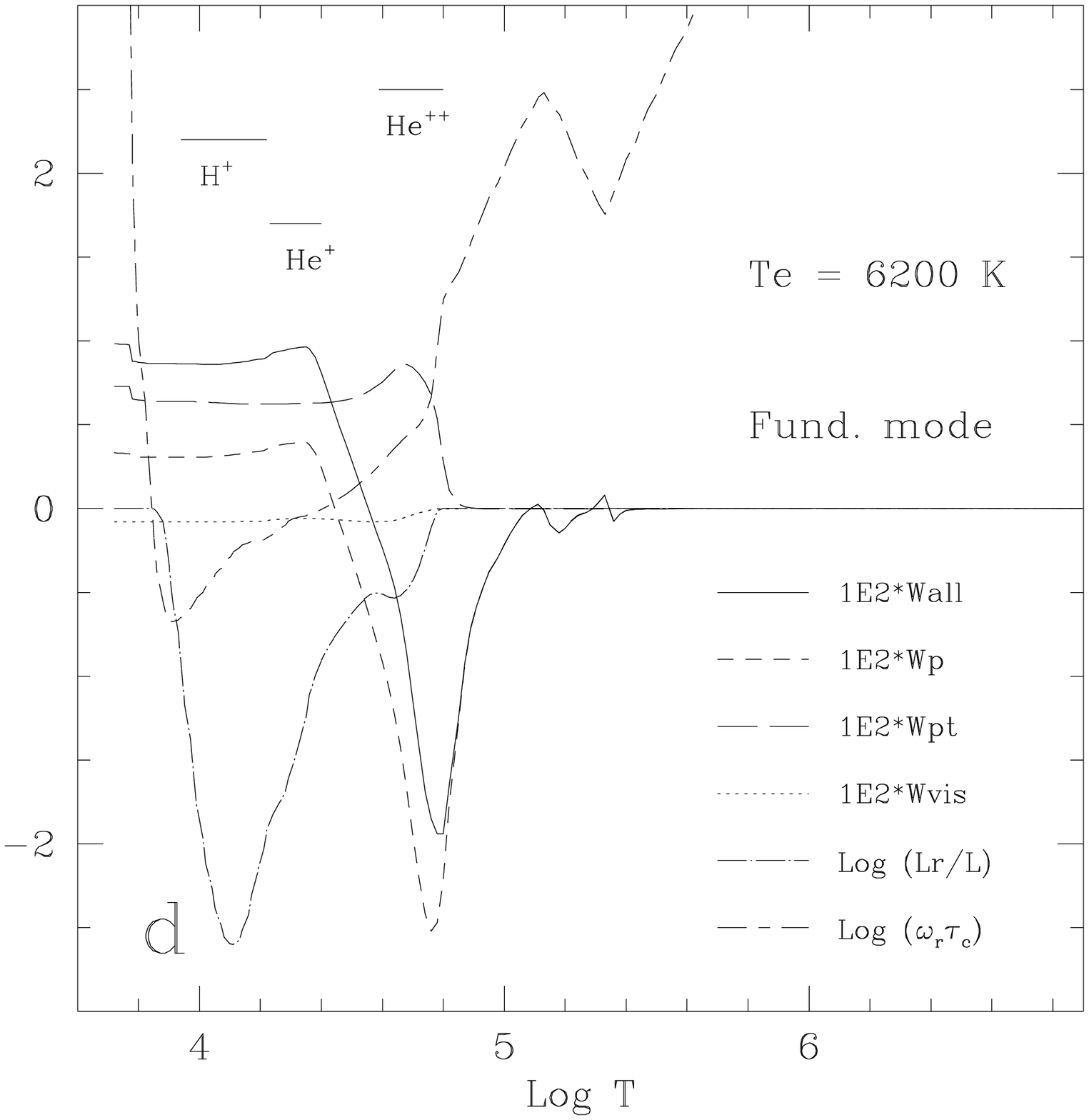,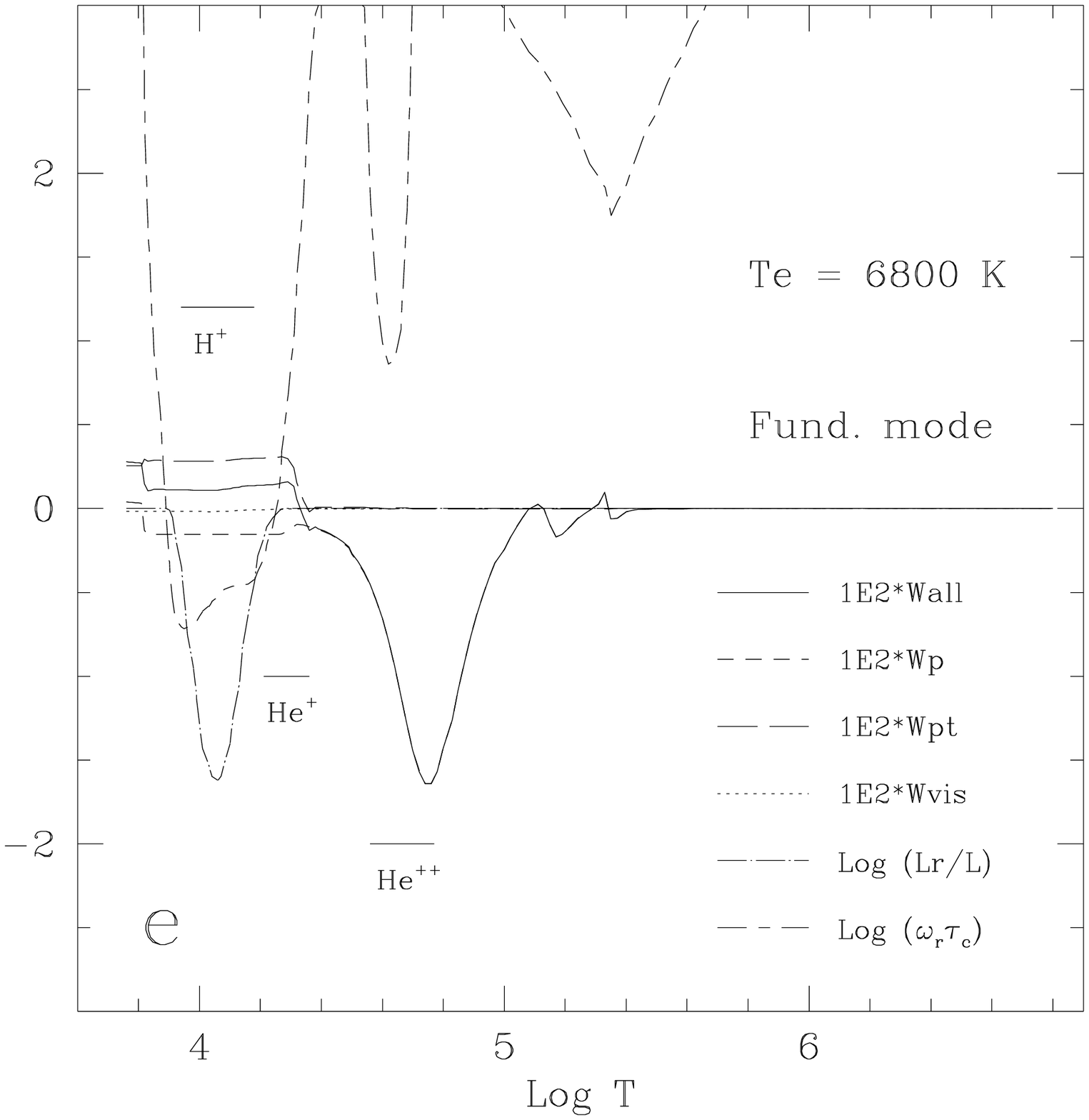,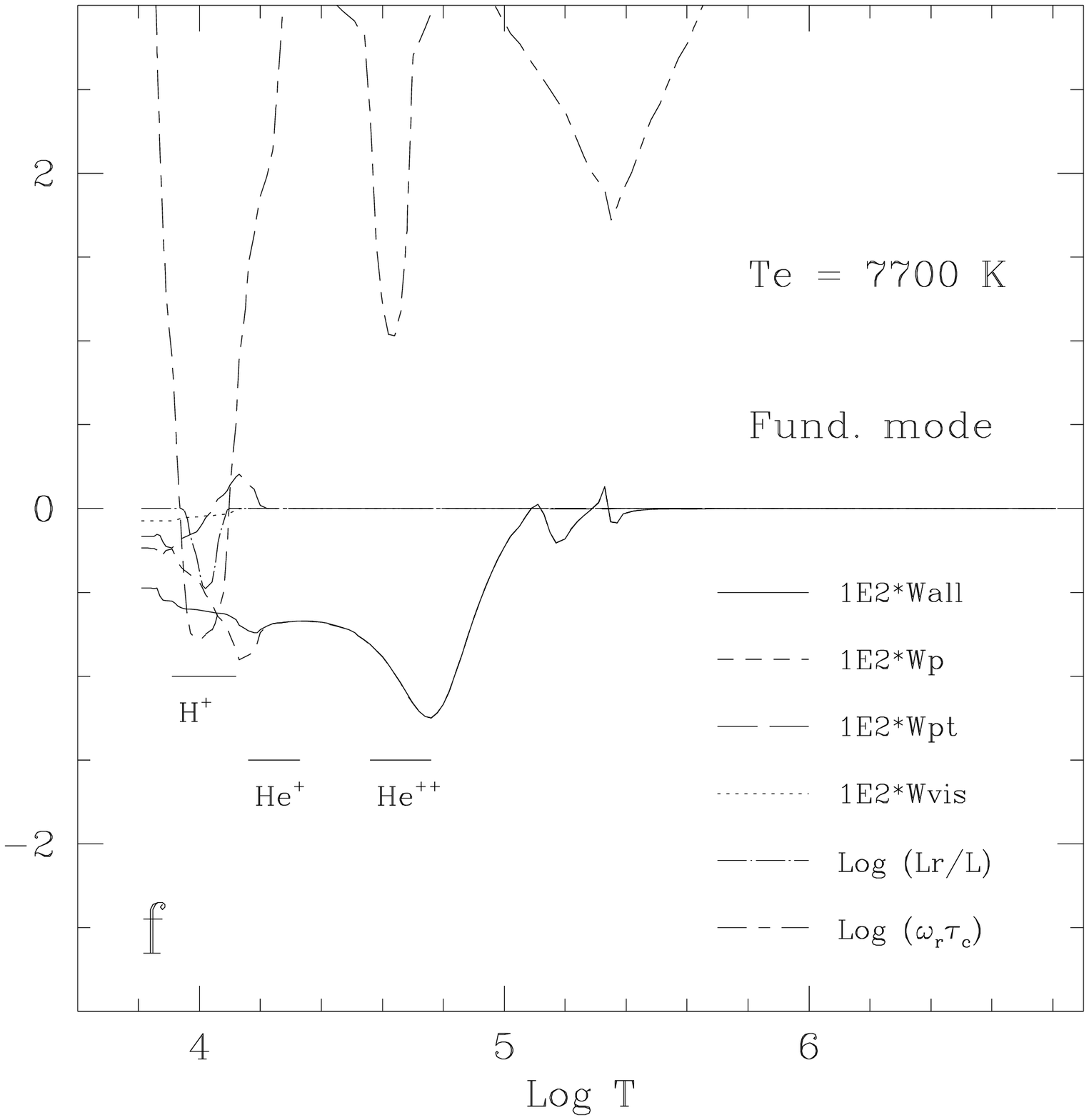]{
The same as Fig.~\ref{fig2}, but the coupling between
convection and oscillations is considered. $W_P$, $W_{Pt}$ and $W_{vis}$
are, respectively, the contributions due to gas (and radiation) pressure, 
turbulent pressure and turbulent viscosity, while $W_{all}$ is the
total integrated work, the other symbols are the same as in Fig.~\ref{fig2}.
\label{fig4}}

\figcaption[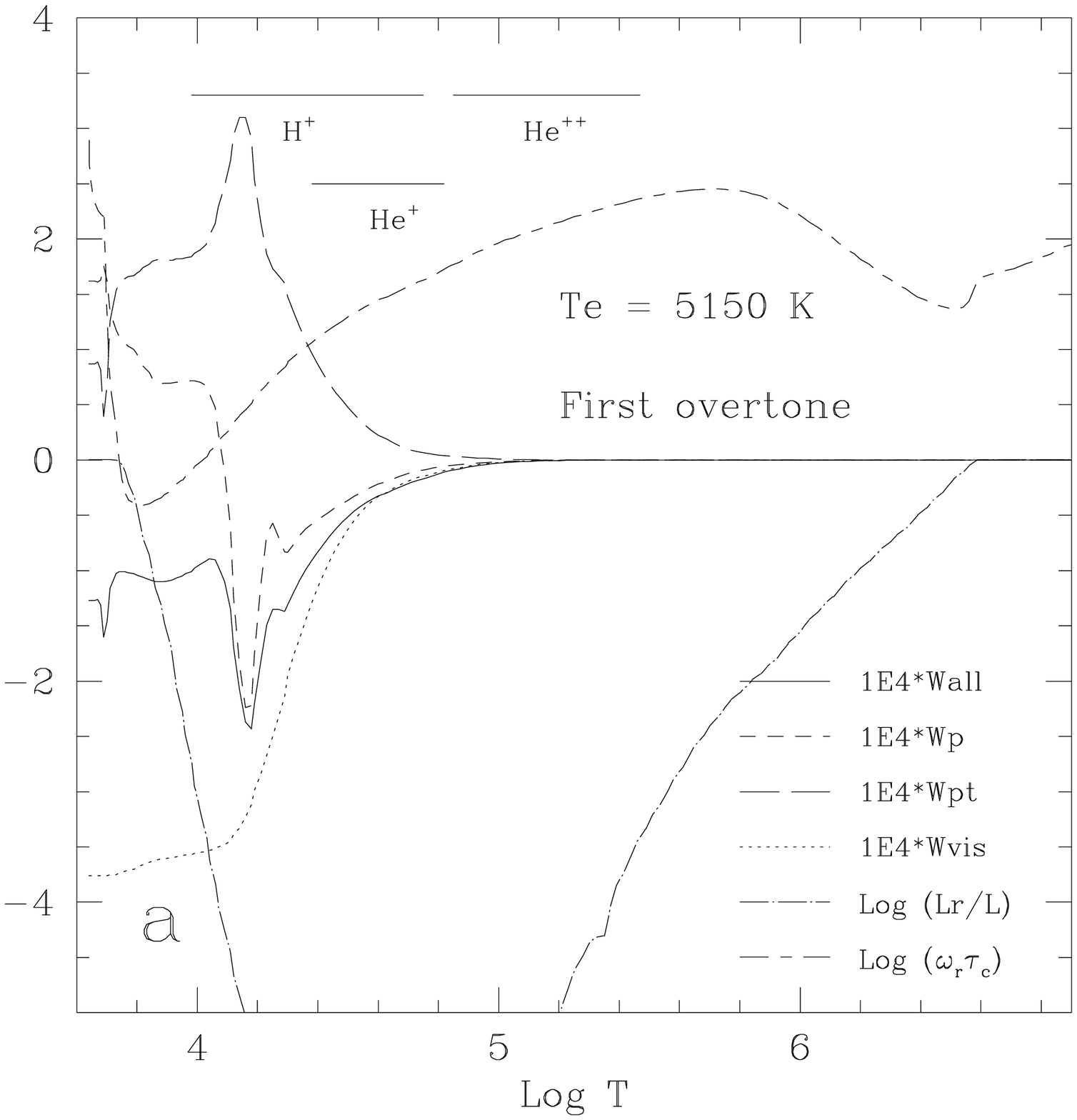,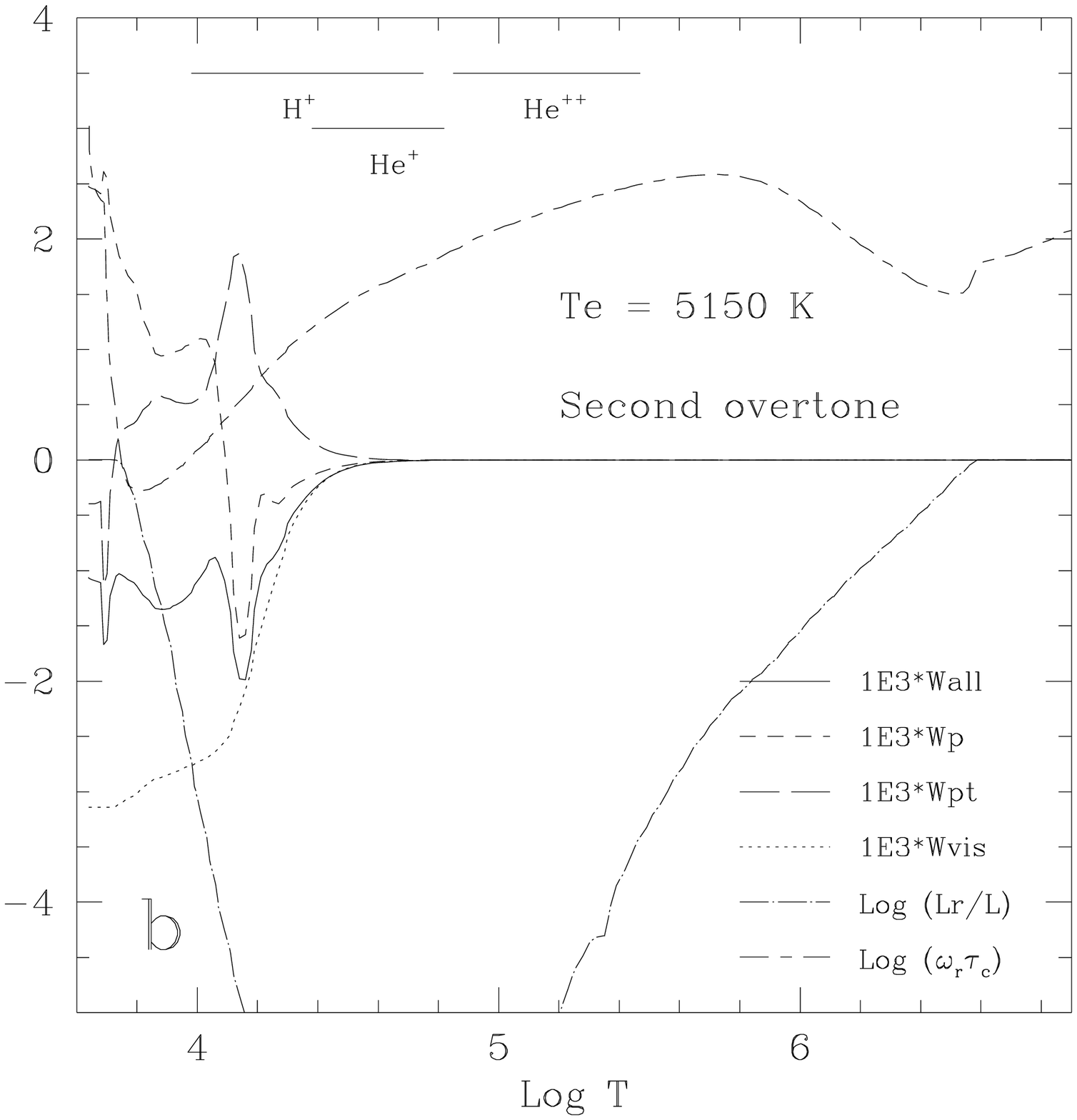,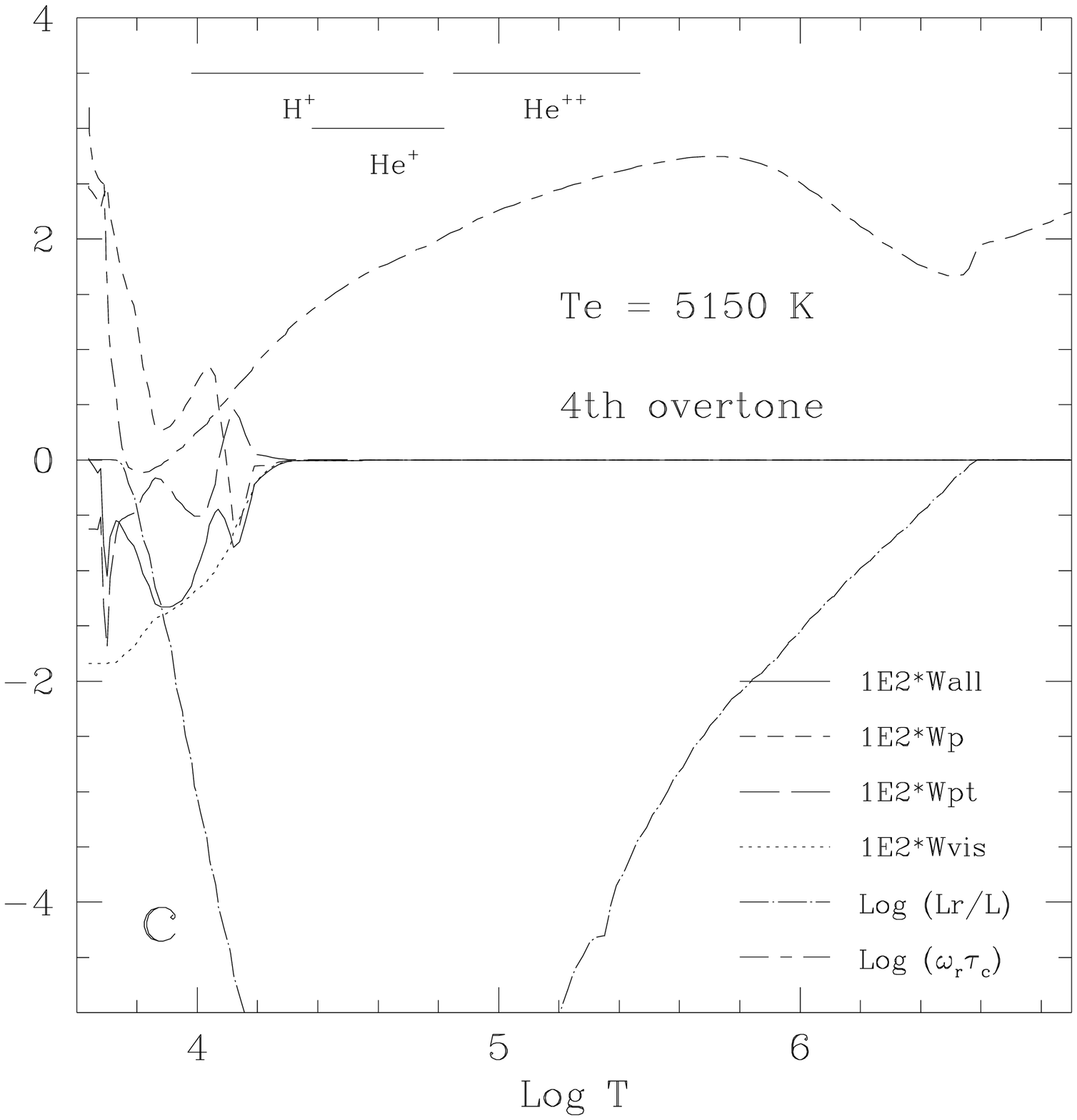]{
The integrated work virsus depth for the first, second, and fourth 
overtones of a $T_e=5150 K$ HB star model, the coupling between
convection and oscillations is considered. Refer to Fig.~\ref{fig4} for
details.
\label{fig5}}

\figcaption[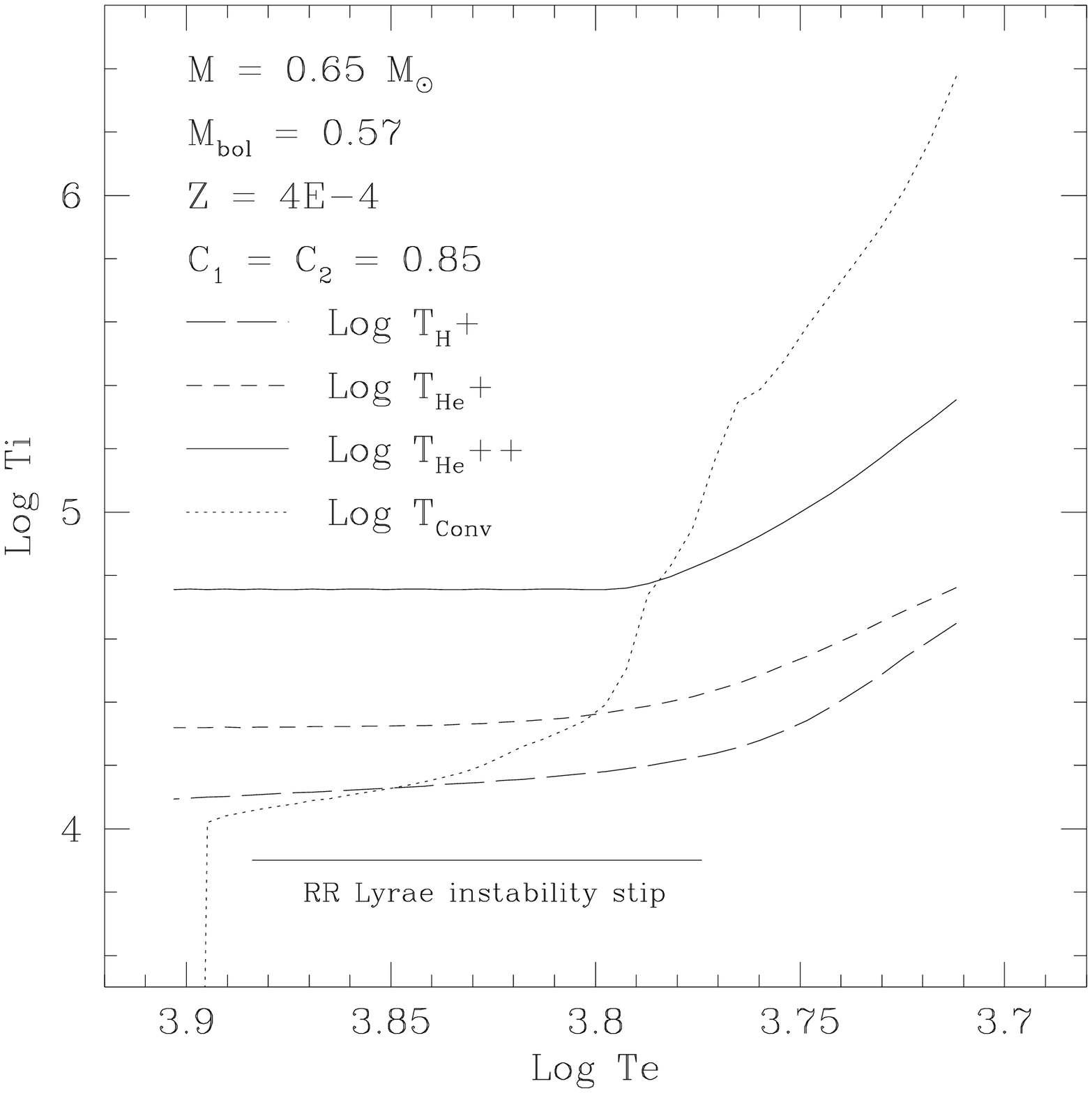]{
The bottom temperatures of the hydrogen, the first and the second helium 
ionization regions and convective region versus effective temperature
for the model series 2 of HB stars. The horizontal lines are the location
of the RR Lyrae instability strip. The horizontal line indicates the 
locations of the RR Lyrae instability sttip.
\label{fig6}}

\figcaption[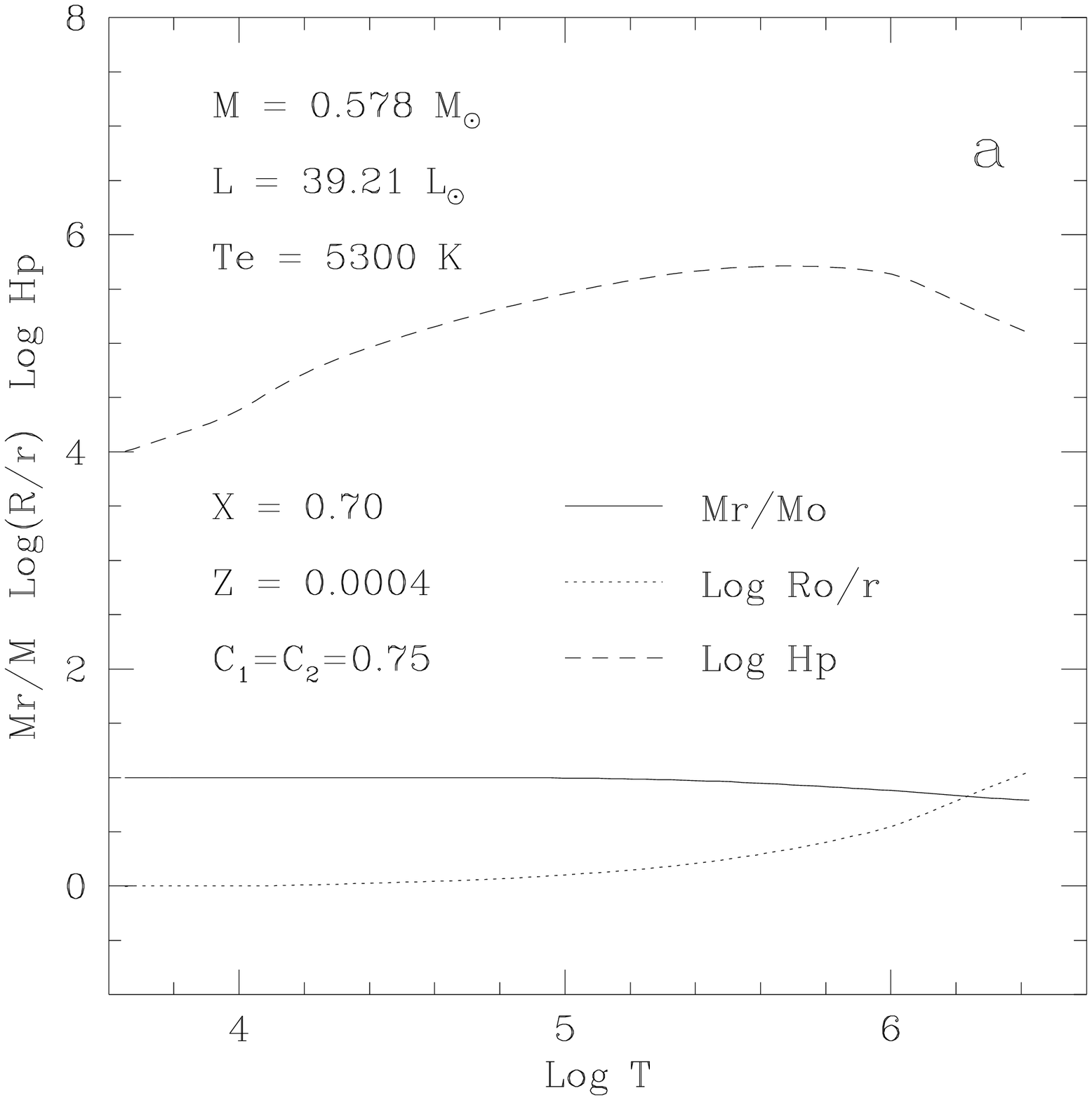,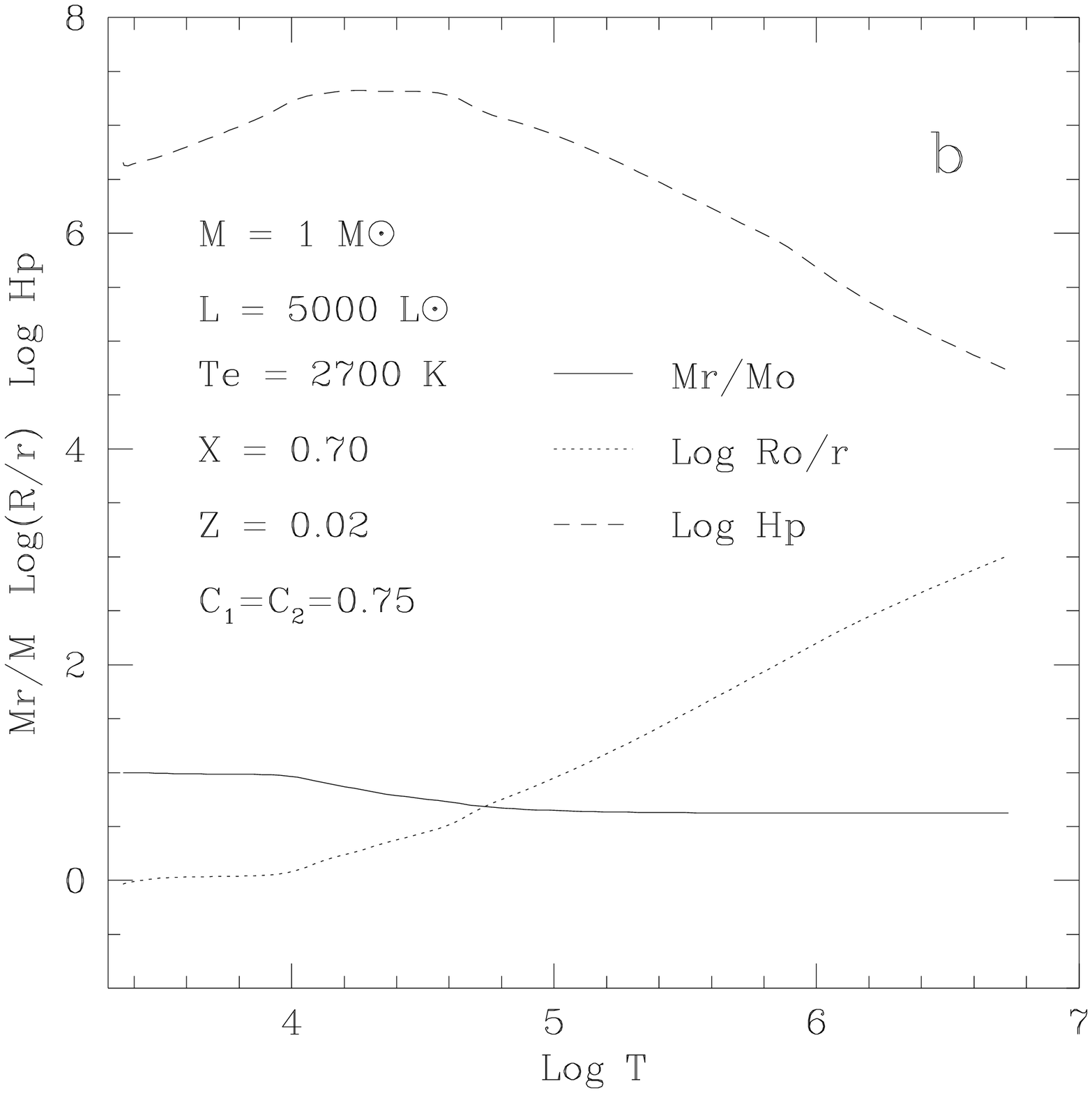]{
The variation of radius $r$, mass $M_r$ and the local
pressure scale height $H_P$ versus depth ($\log T$). Panel a.:
the envelope model of a HB star with $T_e=5300 K$; Panel b.:
a long period variable envelope model with $M=1M_\odot$,
$L=5000L_\odot$, $T_e=2700K$, $x=0.70$ and $z=0.020$.
\label{fig7}}

\newpage
\setcounter{figure}{0}
\begin{figure}
\centerline{\psfig{figure=fig1a.ps,width=7cm}\psfig{figure=fig1b.ps,width=7cm}}
\centerline{\psfig{figure=fig1c.ps,width=7cm}\psfig{figure=fig1d.ps,width=7cm}}
\centerline{\psfig{figure=fig1e.ps,width=7cm}\psfig{figure=fig1f.ps,width=7cm}}
\caption{ }
\end{figure}

\begin{figure}
\centerline{\psfig{figure=fig2a.ps,width=7cm}\psfig{figure=fig2b.ps,width=7cm}}
\centerline{\psfig{figure=fig2c.ps,width=7cm}\psfig{figure=fig2d.ps,width=7cm}}
\centerline{\psfig{figure=fig2e.ps,width=7cm}\psfig{figure=fig2f.ps,width=7cm}}
\caption{ }
\end{figure}

\begin{figure}
\centerline{\psfig{figure=fig3a.ps,width=7cm}}
\centerline{\psfig{figure=fig3b.ps,width=7cm}}
\centerline{\psfig{figure=fig3c.ps,width=7cm}}
\caption{ }
\end{figure}

\begin{figure}
\centerline{\psfig{figure=fig4a.ps,width=7cm}\psfig{figure=fig4b.ps,width=7cm}}
\centerline{\psfig{figure=fig4c.ps,width=7cm}\psfig{figure=fig4d.ps,width=7cm}}
\centerline{\psfig{figure=fig4e.ps,width=7cm}\psfig{figure=fig4f.ps,width=7cm}}
\caption{ }
\end{figure}

\begin{figure}
\centerline{\psfig{figure=fig5a.ps,width=7cm}}
\centerline{\psfig{figure=fig5b.ps,width=7cm}}
\centerline{\psfig{figure=fig5c.ps,width=7cm}}
\caption{ }
\end{figure}

\begin{figure}
\centerline{\psfig{figure=fig6.ps,width=15cm}}
\caption{ }
\end{figure}

\begin{figure}
\centerline{\psfig{figure=fig7a.ps,width=8cm}}
\centerline{\psfig{figure=fig7b.ps,width=8cm}}
\caption{ }

\end{figure}
\begin{figure}
\centerline{\psfig{figure=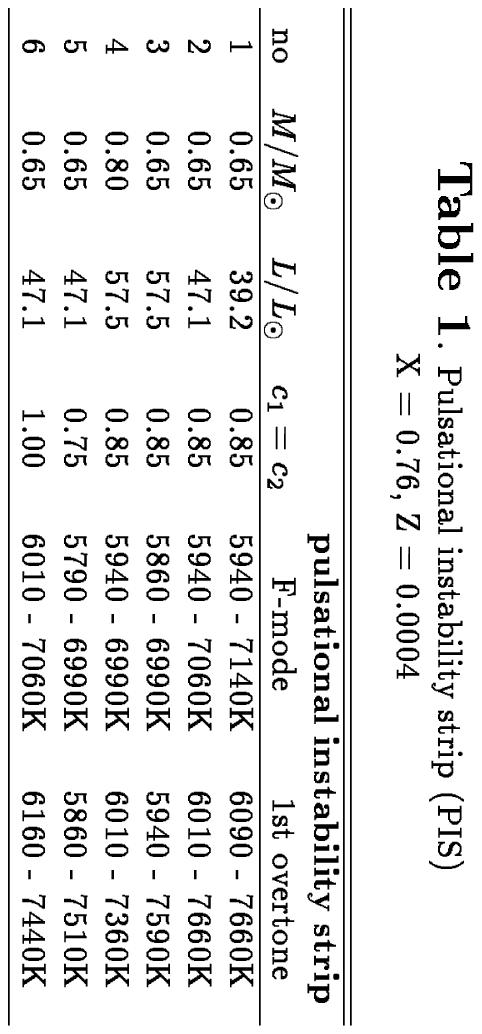}}
\end{figure}
\begin{figure}
\centerline{\psfig{figure=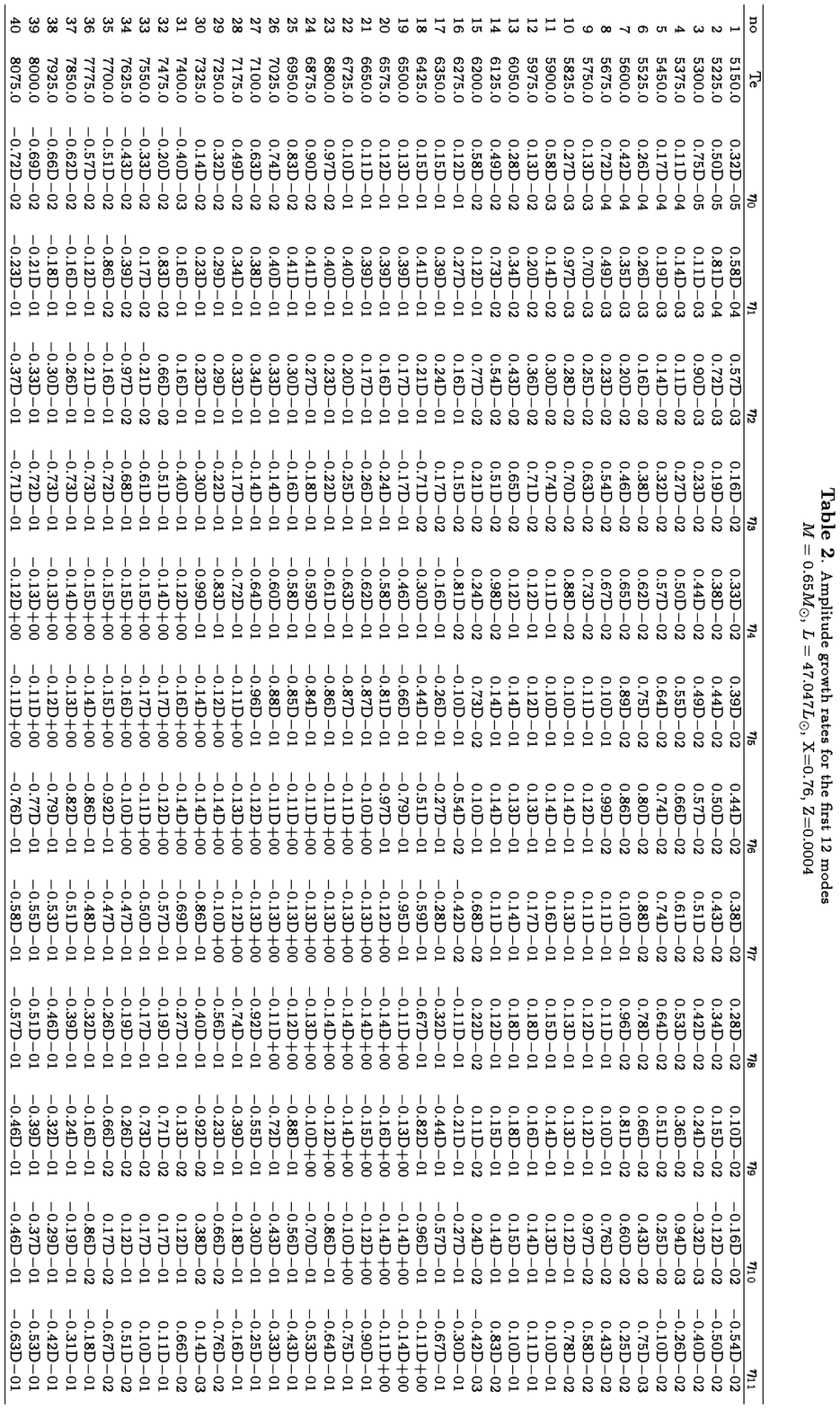}}
\end{figure}
\begin{figure}
\centerline{\psfig{figure=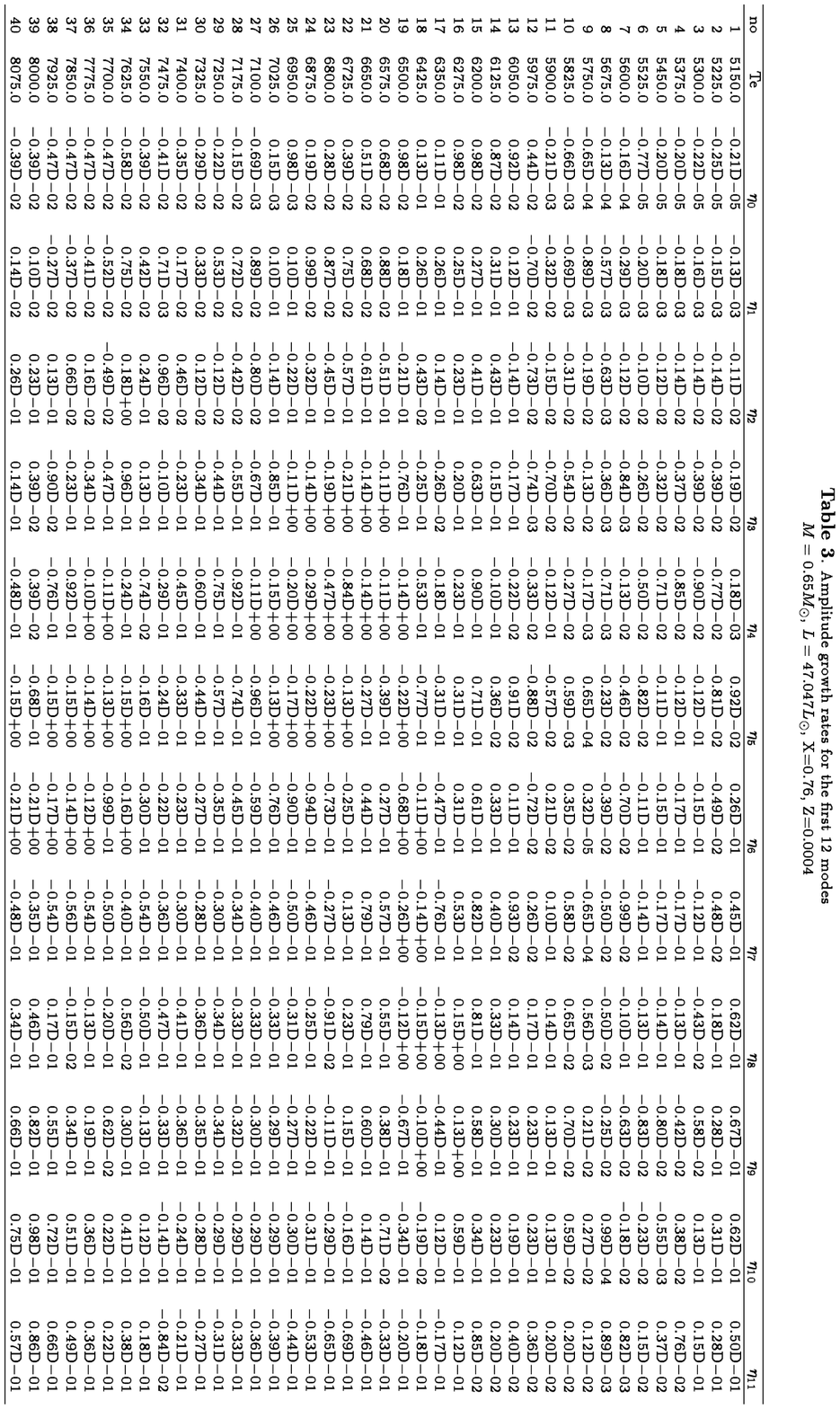}}
\end{figure}
\end{document}